\documentclass[Conference,a4paper]{IEEEtran}
\IEEEoverridecommandlockouts
\usepackage{color}
\usepackage{graphicx}
\usepackage{epstopdf}
\usepackage{amsmath}
\usepackage{amssymb}
\usepackage{algorithm}
\usepackage{algorithmic}
\usepackage{amsmath}
\usepackage{multirow}
\usepackage{booktabs}
\usepackage{array}
\usepackage{amsthm}
\usepackage{stfloats}
\usepackage{caption}
\usepackage{subfigure}
\usepackage{bm}
\usepackage{booktabs}
\usepackage{setspace}
\usepackage{diagbox}
\usepackage{enumerate}
\usepackage{ulem}
\usepackage{url}
\usepackage{circledsteps}

\usepackage{hyperref}
\allowdisplaybreaks[4]

{\bgroup
 \addtolength\abovedisplayshortskip{#1}
 \addtolength\abovedisplayskip{#1}
 \addtolength\belowdisplayshortskip{#1}
 \addtolength\belowdisplayskip{#1}
 }
{\egroup\ignorespacesafterend}

\newcommand{\be}{\begin{equation}}
\newcommand{\ee}{\end{equation}}
\newcommand{\bea}{\begin{eqnarray}}
\newcommand{\eea}{\end{eqnarray}}
\newcommand{\ba}{\begin{array}}
\newcommand{\ea}{\end{array}}

\title{
Max-Min Rate Optimization for Multigroup Multicast MISO Systems Via Novel Transmissive RIS Transceiver
}
\author{\IEEEauthorblockN{Yuan Guo, Wen Chen, Qingqing Wu, Yanze Zhu, Yang Liu, Zhendong Li, and Ying Wang
\thanks{
Y. Guo, W. Chen, Q. Wu, and Z. Zhu are with Department of Electronic Engineering, Shanghai Jiao Tong University, Shanghai, China, 
email:
yuanguo26@sjtu.edu.cn,
wenchen@sjtu.edu.cn,
qingqingwu@sjtu.edu.cn,
yanzezhu@sjtu.edu.cn.
}
\thanks{
Y. Liu  is with the School of Information
and Communication Engineering, Dalian University of Technology, Dalian, China, 
email:
yangliu\_613@dlut.edu.cn.}
\thanks{Z. Li is with the School of Information and Communication Engineering, Xi'an Jiaotong University, Xi'an, China,
email:
lizhendong@xjtu.edu.cn.}
\thanks{Y. Wang is with the State Key Laboratory of Networking and Switching Technology, Beijing University of Posts and Telecommunications, Beijing, China,
email:
wangying@bupt.edu.cn.}
}
}

\begin{document}
\maketitle
\pagestyle{empty}
\thispagestyle{empty}

\begin{abstract}
This paper investigates a novel transmissive
reconfigurable intelligent surface (RIS) transceiver architecture-enabled multigroup multicast downlink communication system.
Under this setup,
an optimization problem is formulated to maximize the minimum rate of users across all groups,
subject to the maximum available power of each RIS unit.
Due to the non-differentiable nature of the objective function,
the max-min rate problem is challenging to solve.
To tackle this difficult problem,
we develop an iterative solution by leveraging the successive convex approximation (SCA) and the penalty function method.
However, 
the above approach has high computational complexity and may lead to compromised performance.
To overcome these drawbacks,
we design an efficient second-order cone programming (SOCP)-based method using the weighted minimum mean squared error (WMMSE) framework to reduce computational complexity.
Furthermore,
to further reduce the computational complexity,
we also propose a low-complexity and solver-free algorithm
that analytically updates all variables 
by combining the smooth approximation theory and the majorization-minimization (MM) method.
Numerical results are provided to verify the convergence and effectiveness of our proposed three algorithms.
It is also demonstrated that
the SOCP-based method outperforms the penalty-based algorithm in terms of both the achieved min rate and the computational complexity.
In contrast, the low-complexity design achieves significantly lower complexity with only slightly degraded performance.

\end{abstract}

\begin{IEEEkeywords}
Transmissive reconfigurable intelligent surface (RIS) transceiver,
multigroup multicast,
max-min rate,
low-complexity algorithm.
\end{IEEEkeywords}

\maketitle
\section{Introduction}
Recently,
among the various promising candidate technologies for 6G,
the rising technology of reconfigurable intelligent surface (RIS) \cite{ref_RIS_1},
which is also widely referred to as 
intelligent reflecting surface (IRS) \cite{ref_RIS_2}
and/or
intelligent surface (IS) \cite{ref_RIS_3},
has been gained great attentions from academia and industry 
and 
is envisioned as a potential solution for the next generation communication system
due to its unique advantages.

In general,
the RIS is a planar surface consisting of a large number of tunable elements, 
which can be realized by varactors and/or positive intrinsic negative (PIN) diodes.
And the tunable unit can independently change the phase shift and/or amplitude of the incident signal.
Since its inherent adaptability,
the flexible deployment of IRS in complex environments
can result in notable enhancements in wireless propagation.
Since RIS is a passive device, 
it enables communication networks to operate cost-effectively with minimal energy and hardware requirements.

Due to the aforementioned merits of the RIS architecture,
a rich body of literature has studied deploying the RIS in wireless systems 
from various perspectives to enhance system performance, e.g., 
\cite{ref_RIS_A_1}$-$\cite{ref_RIS_A_14}.
For instance,
the authors in \cite{ref_RIS_A_1}
considered 
the problem of weighted sum-rate maximization
in RIS-aided  multi-cell communication systems,
aiming to enhance downlink communication for cell-edge users while mitigating inter-cell interference.
The paper \cite{ref_RIS_A_2}
proposed a low-complexity beamforming algorithm to maximize the sum-rate of all multicast groups, 
and the numerical results verified the effectiveness of RIS in downlink multigroup multicast communication systems.
The work \cite{ref_RIS_A_3}
investigated the sum-rate maximization problem over all subcarriers in the RIS-aided
orthogonal frequency division multiplexing (OFDM) system.
The deployment of RIS in millimeter wave (mmWave) multiuser multiple-input multiple-output (MU-MIMO) system was considered in \cite{ref_RIS_A_4},
and the numerical results demonstrated that  RIS can significantly reduce the sum-mean-square-error (sum-MSE).
In secure multiuser communication systems, 
\cite{ref_RIS_A_5}
employed RIS to maximize the weighted minimum approximate ergodic secrecy rate 
under hardware impairments (HIs) at both the RIS and the transceivers.
Furthermore,
in \cite{ref_RIS_A_6},
the authors considered the weighted sum of transferred power maximization
while constraining the secrecy rate
in the RIS-aided secure simultaneous wireless information and power transfer (SWIPT) communication network.
The authors of \cite{ref_RIS_A_7} utilized RIS mounted on an unmanned aerial vehicle (UAV)
to improve communication performance in a  wireless downlink MIMO system.
The paper \cite{ref_RIS_A_8}
designed a near-field wideband beamforming scheme  to maximize system spectral efficiency
and mitigate the double beam split effect in an RIS-aided MIMO system.
The work \cite{ref_RIS_A_9}
adopted a space-time beamforming design to simultaneously improve both the sensing resolution
and accuracy in RIS-empowered multi-target sensing system.
The authors in \cite{ref_RIS_A_10}
employed the RIS in the full-duplex (FD) integrated sensing and communication (ISAC) system to 
improve radar detection probability via suppressing self-interference, 
and developed a low-complexity solution 
that updates all variables analytically and runs highly efficiently.
A novel intelligent omni surface (IOS)-aided ISAC system for the multi-target and multi-user scenario was investigated in \cite{ref_RIS_A_11}.
The study aims to maximize the minimum sensing signal-to-interference-plus-noise ratio (SINR) while ensuring satisfactory communication performance.
The work \cite{ref_RIS_A_12} focused on enhancing the sum-rate while maintaining sensing quality 
in the ISAC system enabled by the novel beyond-diagonal (BD)-RIS architecture.
Furthermore,
based on the previous work \cite{ref_RIS_A_12},
the literature \cite{ref_RIS_A_13} studied the transmit power minimization problem under both the communication and sensing quality constraints in
the BD-RIS-assisted ISAC system.
The authors in \cite{ref_RIS_A_14}
proposed deploying the RIS in the multi-cell ISAC system to minimize the transmit power 
while guaranteeing both communication and sensing requirements.

In addition to the conventional RIS employed as an auxiliary component in wireless networks,
a novel \textit{ transmissive RIS transceiver (TRTC)} architecture,
which is able to achieve greater system performance improvements while consuming less power,
 was introduced in \cite{ref_TRIS_1}.
Differing from conventional multi-antenna systems that rely on active components, 
the proposed TRTC integrates a passive transmissive RIS and a single horn antenna feed,
eliminating the need for numerous RF chains and complex signal processing modules.
Moreover,
compared with the reflective RIS transmitter presented in \cite{ref_RRIS_1}$-$\cite{ref_RRIS_2},
the TRTC technique can effectively solve the following two main problems:
\textit{1) feed source blockage}:
When both the horn antenna and the user are positioned on the same side of the RIS, 
the incident electromagnetic (EM) wave at the reflective RIS transmitter suffers from feed source blockage. 
In contrast, 
as the horn antenna and the user are located on opposite sides of the RIS, 
the TRTC effectively avoids this issue;
\textit{2) echo interference}:
Due to the fact that both the incident and reflected EM waves are located on the same side of the RIS,
the reflective RIS transceiver is susceptible to echo interference.
In contrast, 
the TRTC mitigates this issue by spatially separating the incident and transmitted waves onto opposite sides of the RIS.
Therefore,
the TRTC represents an emerging technology that facilitates sustainable capacity growth in a cost-effective way.

Owing to the advantages of the TRTC architecture, 
recent studies have investigated TRTC-assisted wireless networks from various perspectives to enhance overall system performance,
e.g., \cite{ref_TRIS_A_1}$-$\cite{ref_TRIS_A_10}.
For instance,
a TRTC-aided multi-stream downlink communication system based on time-modulated array (TMA) technology was proposed in \cite{ref_TRIS_A_1},
aiming to maximize the minimum SINR and offering a linear-complexity solution.
In \cite{ref_TRIS_A_2},
the TRTC was employed as a receiver architecture  in the uplink communication system, 
where uplink users adopt the orthogonal frequency division multiple access (OFDMA) technique.
Moreover, 
\cite{ref_TRIS_A_2}
investigated the problem of maximizing the sum-rate of uplink users, subject to quality-of-service (QoS) constraints.
The authors of \cite{ref_TRIS_A_3} studied the sum-rate maximization problem in
the TRTC-assisted SWIPT networks, 
and the simulation results validated that the proposed algorithm can achieve better quality.
In \cite{ref_TRIS_A_4}, 
a TRTC-aided multi-tier computing network architecture was investigated, 
with the objective of minimizing total energy consumption under both communication and computing resource constraints.
The paper \cite{ref_TRIS_A_5} proposed an innovative hybrid active-passive TRTC architecture,
in which each RIS element can dynamically switch between active and passive modes.
Numerical results demonstrated that this flexible design,
applied to a downlink multi-user communication system,
can significantly improve the system's energy efficiency (EE).
The work \cite{ref_TRIS_A_6}
adopted the TRTC to facilitate multi-beam transmission and  directional beam suppression by formulating a max-min metric with non-linear constraints.
Moreover,
to bridge explicit beamforming design with practical implementation,
a realistic model was first presented to accurately capture the input and/or output behavior of the TRTC.
The authors of \cite{ref_TRIS_A_6_1} simultaneously adopted TRTC and RIS in a secure communication system 
and showed that they could significantly boost the weighted sum secrecy rate.
The literature \cite{ref_TRIS_A_7}
designed a time-division sensing communication mechanism in a TRTC-aided robust and secure ISAC system.
Furthermore,
to effectively manage interference and improve resistance to eavesdropping, 
the authors incorporated rate-splitting multiple access (RSMA) as a key enabling technology.
A distributed cooperative ISAC network assisted by the TRTC for enhancing service coverage 
was researched in \cite{ref_TRIS_A_8}.
The study aimed to maximize the minimum radar mutual information (RMI) as the primary performance metric.
The authors of  \cite{ref_TRIS_A_9}  considered maximizing the sum-rate of the multi-cluster
in a Low Earth Orbit (LEO) satellite nonorthogonal multiple access (NOMA) system via using the TRTC architecture.
The paper \cite{ref_TRIS_A_10} applied the TRTC into human activity recognition (HAR).

Nevertheless,
existing researches \cite{ref_TRIS_A_1}$-$\cite{ref_TRIS_A_10} 
have focused solely on evaluating the performance benefits of employing the TRTC architecture under the unicast transmission setup,
where the transmitter delivers a dedicated data stream to each user.
However, 
when the user density is high, 
unicast transmission incurs considerable interference and system overhead.
In contrast, 
multicast transmission, 
in which a common data stream is simultaneously delivered to multiple users, 
offers a highly efficient solution for broadcasting shared information in practical scenarios such as live video streaming and online gaming.
Therefore, 
exploring the potential of TRTC in multicast transmission is of great significance.

Inspired by the above inspections,
we are motivated to enhance the performance of the multigroup multicast communication systems by employing TRTC devices characterized by low-cost and low-power consumption.
Towards this end,
this paper considers a TRTC-aided multigroup multicast downlink communication system.
Specifically,  
the contributions of this paper are elaborated as follows:
\begin{itemize}
\item
This paper considers the beamforming design in a multigroup multicast multiple-input single-output (MISO) downlink communication system empowered by the novel TRTC architecture,
aiming to explore potential performance gains.
We investigate the problem of maximizing the sum-rate across all multicast groups, 
where the rate of each group is constrained by the minimum rate among its users, 
subject to the individual maximum transmit power limits of each TRTC element.
To the best of our knowledge, 
this problem has not been explored in the existing literature, e.g., \cite{ref_TRIS_A_1}$-$\cite{ref_TRIS_A_10}.

\item
Due to the complex and non-differentiable nature of the objective function, 
solving the highly non-convex max-min rate problem is particularly challenging. 
To tackle this optimization problem, 
we first convert the non-differentiable concave objective function into a set of constraints by introducing slack variables, 
and then equivalently reformulate it as a rank-constrained semidefinite programming (SDP) problem. 
By integrating the successive convex approximation (SCA)  technique \cite{ref_MM} with the penalty function method, 
we successfully develop an iterative algorithm to solve the resulting SDP problem.

\item
To reduce computational complexity, 
we reformulate the original problem as a second-order cone programming (SOCP) problem 
via combining the weighted minimum mean squared error (WMMSE) framework \cite{ref_WMMSE} with the introduction of slack variables. 
Based on this reformulation, 
we design an efficient iterative algorithm to solve the SOCP problem.

\item
Furthermore, 
we also develop a low-complexity solution that avoids reliance on any numerical solver, e.g., CVX.
Specifically,
based on the block diagonal structure of the quadratic term coefficient and the TRTC element power constraint,
we decompose the original variable into multiple subvariables.
And then,
the non-differentiable concave objective function is first approximated by a differentiable one using smooth approximation theory \cite{ref_Log_sum}.
Subsequently,
by exploiting the majorization-minimization (MM) method \cite{ref_MM} and analyzing optimality conditions, 
we are able to derive analytical solutions for all variables.

\item
Last but not least,
extensive numerical results 
are presented to validate
the effectiveness and efficiency of our proposed three solutions.
The results demonstrate that the SOCP-based algorithm outperforms the penalty-based approach in terms of both the achieved group sum-rate 
and computational complexity. 
Meanwhile, 
the MM-based design exhibits significantly lower computational complexity, with only a slight performance trade-off.

\end{itemize}

The rest of the paper is organized as follows. 
Section II will introduce the model of the TRTC-enabled multigroup multicast communication system
 and formulate the max-min rate optimization problem.
Sections III and IV will propose the penalty-based and SOCP-based solutions to tackle the proposed problem, respectively. 
A low-complexity algorithm will be developed in Section V. 
Sections VI and VII will present numerical results and conclusions of the paper, respectively.

\textit{Notations:}
Lower-case and boldface capital letters are respectively represented as 
vectors and matrices;
$\mathbf{X}^{\ast}$,
$\mathbf{X}^{T}$,
and
$\mathbf{X}^{H}$
denote the conjugate, transpose, and
conjugate transpose of matrix $\mathbf{X}$, respectively;
$\mathbb{C}^{N \times 1}$ represents the set of $N \times 1$ complex vectors;
$\mathbf{0} $ denotes the all zeros matrix;
$\Vert \mathbf{x} \Vert_{2}$ denotes the $l_2$ norm of the vector $\mathbf{x}$;
$\Vert \mathbf{X} \Vert_{2}$,
$\Vert \mathbf{X} \Vert_{F}$,
and
$\Vert \mathbf{X} \Vert_{\ast}$
stand for
spectral norm,
Frobenius norm,
and nuclear norm of matrix $\mathbf{X}$, respectively; 
The largest eigenvalue of matrix $\mathbf{X}$
and the corresponding eigenvector 
are denoted by $\lambda_{\text{max}}(\mathbf{X})$ and $
\boldsymbol{\lambda}_{\text{max}}(\mathbf{X})$, 
respectively;
$ \triangleq $ and $ \sim $ signify ``defined as''
and ``distributed as'', respectively;
$\text{Tr}\{ \mathbf{X} \}$
and 
$\text{Rank}(\mathbf{X})$
 represent the trace and rank of matrix $\mathbf{X}$, respectively;
$\mathbf{X} \succeq \mathbf{0} $ indicates that  
$\mathbf{X} $ is a positive semidefinite (PSD) matrix;
$ \mathbb{E}[\cdot]$ denotes the statistical expectation;
$\mathcal{CN}(\mathbf{x}, \boldsymbol{\Sigma})$
denotes the distribution of a circularly symmetric complex Gaussian (CSCG)
vector with mean vector $\mathbf{x}$
and covariance matrix $\boldsymbol{\Sigma}$;
$\nabla_{ \mathbf{X} } f( \mathbf{X} )$ 
represents the gradient of the real-valued continuous function 
$f(\mathbf{X})$ with respect to matrix $\mathbf{X} $;
$\text{diag}(\mathbf{x})$ denotes a diagonal matrix whose 
diagonal entries are given by the elements of the vector $\mathbf{x}$;
$\text{blkdiag}(\mathbf{X}_1, \cdots, \mathbf{X}_N)$
represents a block diagonal matrix 
with $\mathbf{X}_1, \cdots, \mathbf{X}_N$ as its diagonal blocks.

\section{System Model and Problem Formulation}
\subsection{System Model}

\begin{figure}[t]
	\centering
	\includegraphics[width=.35\textwidth]{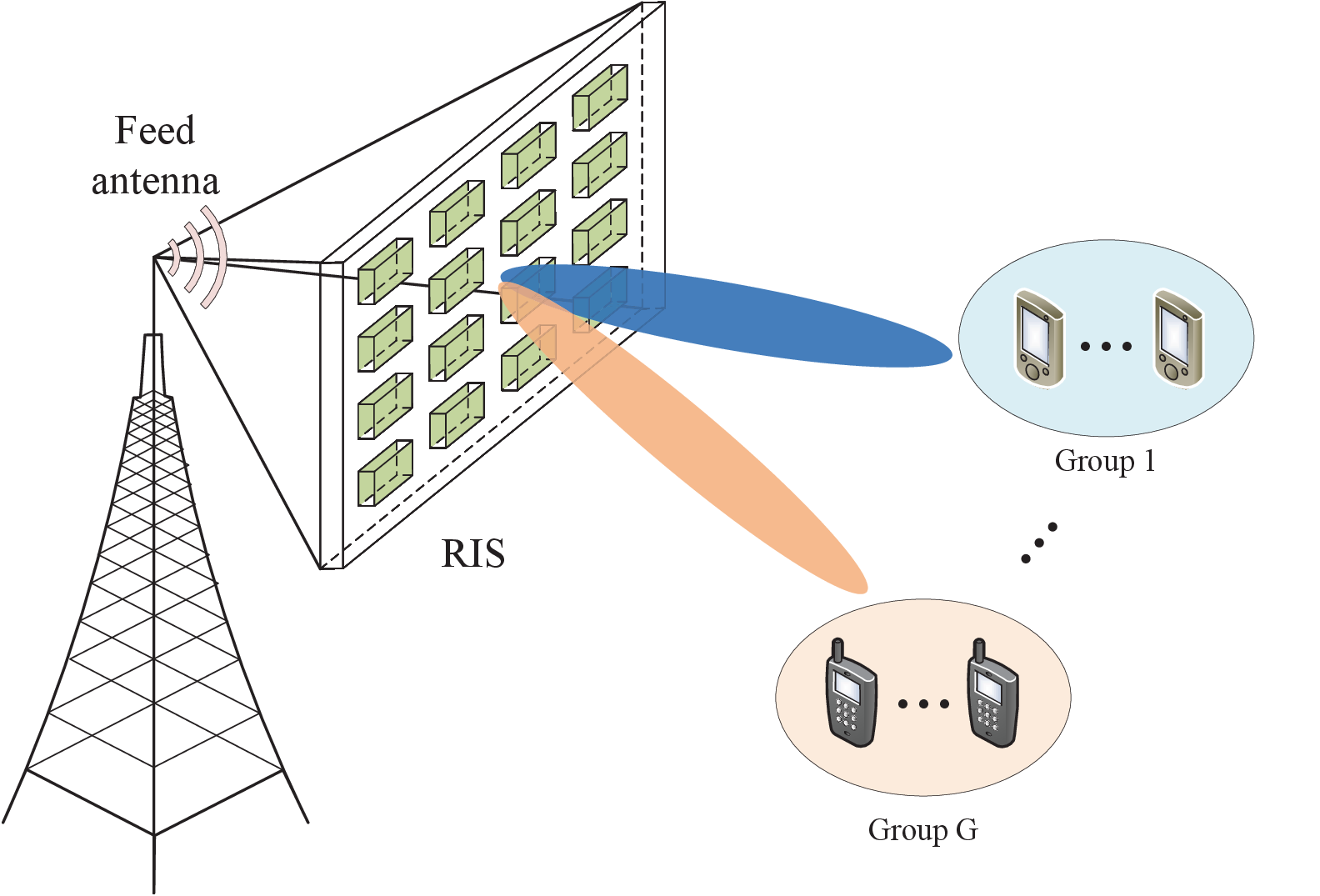}
	\caption{A TRTC enabled multigroup multicast communication  system.}
	\label{fig.1}
\end{figure}

As shown in Fig. \ref{fig.1},
we consider a TRTC-enabled 
multigroup multicast MISO communication system,
where a TRTC
equipped with $N$ elements serves $K$ single-antenna users grouped into $G$ multicast groups.
For convenience, 
the sets of users, multicast groups, and TRTC units are denoted
as 
$\mathcal{K} = \{1,2,\cdots,K\}$,
$\mathcal{G} = \{1,2,\cdots,G\}$,
and
$\mathcal{N} = \{1,2,\cdots,N\}$, respectively.
Let $\mathcal{K}_{g}$ denotes as
the user set belonging to group $g\in \mathcal{G}$.
Besides,
each user can only belong to one group,
i.e., $\mathcal{K}_{i} \cap\mathcal{K}_{j} = \varnothing$, $\forall i,j \in \mathcal{G}$, $i\neq j$.
The transmit signal at the TRTC can be given as
\begin{align}
\mathbf{x} = {\sum}_{g=1}^{G}\mathbf{f}_gs_g, \forall g\in \mathcal{G},
\end{align}
where
$s_g$ denotes the independent Gaussian data symbol of the $g$-th group
and follows $ \mathbb{E}[\vert s_g\vert^2]=1$,
and the vector $\mathbf{f}_g \in \mathbb{C}^{N \times 1}$
represents the corresponding beamformer.

Moreover,
according to the signal generation mechanism of the TRTC \cite{ref_TRIS_1},
the beamforming vectors will  
satisfy the following each TRTC unit power constraint
\begin{align}
 \mathbf{f}^H\mathbf{\bar{A}}_n\mathbf{f} \leq P_t, \forall n \in \mathcal{N},
\end{align}
where
$\mathbf{f} \triangleq [ \mathbf{f}_1^T,\mathbf{f}_2^T,\cdots,\mathbf{f}_G^T ]^T \in \mathbb{C}^{NG\times 1}$,
an index vector $\mathbf{a}_n$ indicates that  the $n$-th position is $1$ and other positions are $0$, i.e., $\mathbf{a}_n \triangleq [0,0,\cdots,\underbrace{1}\limits_{\textrm{n-}th},\cdots,0]^T \in \mathbb{R}^{N \times 1}$,
$\mathbf{A}_n \triangleq \textrm{diag}(\mathbf{a}_n )\in \mathbb{R}^{N\times N}$,
$ \mathbf{\bar{A}}_n \triangleq \text{blkdiag}(  \mathbf{A}_n,\cdots ,\mathbf{A}_n  )\in \mathbb{R}^{NG\times NG} $,
and 
$P_t$ denotes the maximum transmission power for each TRTC unit.

The received signal at the $k$-th user belonging to group $g$ can be represented as
\begin{align}
y_k\! = \!
\mathbf{\bar{h}}_k^H \mathbf{B}_g \mathbf{f}s_g
\!\!+\!\!\! {\sum}_{i\neq g}^{G} \mathbf{\bar{h}}_k^H \mathbf{B}_i\mathbf{f} s_i
\!+\! n_k,
\forall k \in \mathcal{K}_g, 
\forall g \in \mathcal{G}, 
\end{align}
where 
$\mathbf{h}_k \in \mathbb{C}^{N\times 1}$ represents the channel from the TRTC to the $k$-th user
and
$n_k \sim \mathcal{CN}(0,\sigma_k^2)$ 
is the complex additive white Gaussian noise (AWGN) at the $k$-th user,
$\mathbf{\bar{h}}_k \triangleq [\mathbf{{h}}_k^T, \cdots,\mathbf{{h}}_k^T]^T\in\mathbb{C}^{NG\times 1} $,
an index vector $\mathbf{b}_g$ indicates that
the value of the positions within the range $((g-1)\times N +1 )\sim (g\times N)$ is 1 and other positions are $0$,
i.e.,
$\mathbf{b}_g \triangleq [0,\cdots ,0,\underbrace{1,\cdots,1}\limits_{\textrm{N}},0,\cdots,0] \in \mathbb{R}^{NG\times 1}  $,
and
$\mathbf{B}_g \triangleq \text{diag}( \mathbf{b}_g ) \in \mathbb{R}^{NG\times NG}$, $\forall g \in \mathcal{G}$.

The SINR for the $k$-th  user is obtained as
\begin{align}
\textrm{SINR}_k = 
\frac{\vert\mathbf{\bar{h}}_k^H  \mathbf{B}_g \mathbf{f} \vert^2}
{ \sum_{i\neq g}^{G}\vert\mathbf{\bar{h}}_k^H\mathbf{B}_i\mathbf{f}\vert^2 + \sigma_k^2},
\forall k \in \mathcal{K}_g, 
\forall g \in \mathcal{G}, 
\end{align}
and the achievable rate of each user can be written as
\begin{align}
\mathrm{R}_k(\mathbf{f}) = \textrm{log}(1+\textrm{SINR}_k), 
\forall k \in \mathcal{K}_g, 
\forall g \in \mathcal{G}.
\end{align}

\subsection{Problem Formulation}
Due to the characteristics of the multicast communication framework,
our goal is to maximize the minimum rate of each group 
via optimizing the transmit beamformer $\mathbf{f}$.
Mathematically, 
the optimization problem is formulated as
\begin{subequations}
\begin{align}
\textrm{(P0)}:&\mathop{\textrm{max}}
\limits_{\mathbf{f}
}\
\bigg\{ \mathrm{R}_{s} (\mathbf{f}) 
= 
{\sum}_{g=1}^{G} 
\mathop{\textrm{min}}
\limits_{ k \in \mathcal{K}_g } 
\mathrm{R}_k(\mathbf{f}) 
 \bigg\}\label{P0_obj}\\
\textrm{s.t.}\ 
& \mathbf{f}^H\mathbf{\bar{A}}_n\mathbf{f} \leq P_t, \forall n \in \mathcal{N},\label{P0_c_1}
\end{align}
\end{subequations}

The problem (P0) is a non-convex problem and challenging to solve 
since its highly non-differentiable and non-convex objective function.

\section{Penalty-based Method}

In this section,
we will propose a penalty-based method to solve problem (P0).
Firstly,
to make the problem (P0) more tractable,
by introducing the slack variables $\{t_g\}$,
problem (P0) can be equivalently transformed as follows
\begin{subequations}
\begin{align}
\textrm{(P1)}:&\mathop{\textrm{max}}
\limits_{\mathbf{f}, \{ t_g \}
}\
{\sum}_{g=1}^{G} t_g 
 \\
\textrm{s.t.}\ 
&  \mathrm{R}_k(\mathbf{f})  \geq t_g, 
\forall k \in \mathcal{K}_g, 
\forall g \in \mathcal{G}, \\
& \mathbf{f}^H\mathbf{\bar{A}}_n\mathbf{f} \leq P_t, \forall n \in \mathcal{N}.
\end{align}
\end{subequations}

Note that
the optimization problem (P1) is still non-convex with respect to (w.r.t.) the variable $\mathbf{f}$. 
To make the problem more tractable,
by defining \underline{$\mathbf{F} \triangleq \mathbf{f} \mathbf{f}^H \in \mathbb{C}^{NG\times NG} $},
we can equivalently recast the optimization problem (P1) as
a rank-constrained semidefinite programming (SDP) problem, 
which is given as
\begin{subequations}
\begin{align}
\textrm{(P2)}:&\mathop{\textrm{max}}
\limits_{\mathbf{F}, \{ t_g \}
}\
{\sum}_{g=1}^{G} t_g 
\\
\textrm{s.t.}\ 
&  \mathrm{\tilde{R}}_k(\mathbf{F})  \geq t_g, 
\forall k \in \mathcal{K}_g, 
\forall g \in \mathcal{G}, \label{P2_c_1}\\
& \text{Tr}(\mathbf{F}\mathbf{\bar{A}}_n) \leq P_t, \forall n \in \mathcal{N},\\
& \mathbf{F} \succeq \mathbf{0},\\
& \text{Rank}(\mathbf{F}) = 1.\label{P2_c_4}
\end{align}
\end{subequations}
where
\begin{align}
\mathrm{\bar{R}}_k(\mathbf{F}) 
\triangleq
\textrm{log}\bigg(1+\frac{
\textrm{Tr}( \mathbf{B}_g \mathbf{\bar{h}}_k\mathbf{\bar{h}}_k^H \mathbf{B}_g \mathbf{F} ) }
{ \sum_{i\neq g}^{G}
\textrm{Tr}( \mathbf{B}_i \mathbf{\bar{h}}_k\mathbf{\bar{h}}_k^H \mathbf{B}_i \mathbf{F} )
 + \sigma_k^2} \bigg).
\end{align}

However,
the problem (P2) is still non-convex due to the constraints (\ref{P2_c_1}) and (\ref{P2_c_4}).
Next,
we will deal with the constraints (\ref{P2_c_1}) and (\ref{P2_c_4}) one by one.
First,
we rewrite the left of the constraint (\ref{P2_c_1}) in the form of the difference of convex (DC) function,
which can be expressed as
\begin{align}
\mathrm{\bar{R}}_k(\mathbf{F}) 
= \mathrm{\dot{R}}_k(\mathbf{F})  - \mathrm{\ddot{R}}_k(\mathbf{F}),
\end{align}
where
\begin{align}
&\mathrm{\dot{R}}_k(\mathbf{F}) \triangleq  \textrm{log}\big(
 {\sum}_{ g=1}^{G}
\textrm{Tr}( \mathbf{B}_g \mathbf{\bar{h}}_k\mathbf{\bar{h}}_k^H \mathbf{B}_g \mathbf{F} )
 + \sigma_k^2 \big),\\
&\mathrm{\ddot{R}}_k(\mathbf{F})\triangleq
\textrm{log}\big(
 {\sum}_{ i\neq g}^{G}
\textrm{Tr}( \mathbf{B}_g \mathbf{\bar{h}}_k\mathbf{\bar{h}}_k^H \mathbf{B}_g \mathbf{F} )
 + \sigma_k^2 \big).
\end{align}

Since two functions $\mathrm{\dot{R}}_k(\mathbf{F})$
 and
  $\mathrm{\ddot{R}}_k(\mathbf{F})$ 
  are both concave in terms of the variable $\mathbf{F}$,
  the constraint $\mathrm{\dot{R}}_k(\mathbf{F})  - \mathrm{\ddot{R}}_k(\mathbf{F})\geq t_g $ is non-convex.

Following the SCA method \cite{ref_MM},
we linearize the non-convex term $\mathrm{\ddot{R}}_k(\mathbf{F})$ to obtain a tight upper bound as follows
\begin{align}
\mathrm{\ddot{R}}_k(\mathbf{F}) \leq \mathrm{\ddot{R}}_k(\mathbf{F}_0) 
+ \textrm{Tr}( \nabla_{\mathbf{F}}^H\mathrm{\ddot{R}}_k(\mathbf{F}_0)(\mathbf{F}- \mathbf{F}_0) ) ,
\end{align}
where $\mathbf{F}_0$ is the value obtained in the last iteration,
and the gradient of function $\mathrm{\ddot{R}}_k$ w.r.t. $\mathbf{F}$ is given as
\begin{align}
\nabla_{\mathbf{F}}\mathrm{\ddot{R}}_k(\mathbf{F}_0)
\triangleq
\frac{ {\sum}_{ i\neq g}^{G}
( \mathbf{B}_g \mathbf{\bar{h}}_k\mathbf{\bar{h}}_k^H \mathbf{B}_g  )}
{ {\sum}_{ i\neq g}^{G}
\textrm{Tr}( \mathbf{B}_g \mathbf{\bar{h}}_k\mathbf{\bar{h}}_k^H \mathbf{B}_g \mathbf{F} )
 + \sigma_k^2}.
\end{align}

Based on the above transformations,
the problem (P2) can be rewritten as
\begin{subequations}
\begin{align}
\textrm{(P3)}:&\mathop{\textrm{min}}
\limits_{\mathbf{F}, \{ t_g \}
}\
-{\sum}_{g=1}^{G} t_g 
\\
\textrm{s.t.}\ 
&  \mathrm{\dot{R}}_k(\mathbf{F})   
- \textrm{Tr}( \nabla_{\mathbf{F}}^H\mathrm{\ddot{R}}_k(\mathbf{F}_0)(\mathbf{F}- \mathbf{F}_0) )\\
&- \mathrm{\ddot{R}}_k(\mathbf{F}_0)
 \geq t_g, 
\forall k \in \mathcal{K}_g, 
\forall g \in \mathcal{G}, \nonumber\\
& \text{Tr}(\mathbf{F}\mathbf{\bar{A}}_n) \leq P_t, \forall n \in \mathcal{N},\\
& \mathbf{F} \succeq \mathbf{0},\\
& \text{Rank}(\mathbf{F}) = 1.\label{P3_c_4}
\end{align}
\end{subequations}

It is important to note that
the only remaining non-convexity of the problem (P3) arises from the rank constraint (\ref{P3_c_4}).
Next,
we continue to tackle the rank-one constraint (\ref{P3_c_4}),
which can be equivalently written as
\begin{align}
\text{Rank}(\mathbf{F}) = 1
\Leftrightarrow
\Vert\mathbf{F}\Vert_{\ast} - \Vert\mathbf{F}\Vert_{2} \leq 0,
\end{align}
where
$\Vert \cdot \Vert_{2}$
and
$ \Vert \cdot \Vert_{\ast} $
represent the spectral norm and nuclear norm, respectively.
For the any positive semi-definite matrix $\mathbf{F}$,
the following  inequality can be held:
\begin{align}
\Vert\mathbf{F}\Vert_{\ast} = {\sum}_{i} \sigma_{1,i} \geq \Vert\mathbf{F}\Vert_{2}
 = \mathop{\text{max}}\limits_{i} \ \sigma_{1,i}, \label{Rank_one_transformation}
\end{align}
where
$\sigma_{1,i}$ is the $i$-th singular value of the matrix $\mathbf{F}$.
Besides,
when the rank of the matrix $\mathbf{F}$ is one,
the equation  (\ref{Rank_one_transformation}) can achieve equality.

Next,
we propose a penalty-based method \cite{ref_Penalty_1}$-$\cite{ref_Penalty_2},
to tackle it.
By adding the constraint (\ref{Rank_one_transformation}) into the objective function, 
problem (P3) is transformed into
\begin{subequations}
\begin{align}
\textrm{(P4)}:&\mathop{\textrm{min}}
\limits_{\mathbf{F}, \{ t_g \}
}\
-{\sum}_{g=1}^{G} t_g + \frac{1}{2\rho}( \Vert\mathbf{F}\Vert_{\ast} - \Vert\mathbf{F}\Vert_{2} )\label{P4_obj}
\\
\textrm{s.t.}\ 
&  \mathrm{\dot{R}}_k(\mathbf{F})   
- \textrm{Tr}( \nabla_{\mathbf{F}}^H\mathrm{\ddot{R}}_k(\mathbf{F}_0)(\mathbf{F}-\mathbf{F}_0) )\\
&- \mathrm{\ddot{R}}_k(\mathbf{F}_0)
 \geq t_g, 
\forall k \in \mathcal{K}_g, 
\forall g \in \mathcal{G}, \nonumber\\
& \text{Tr}(\mathbf{F}\mathbf{\bar{A}}_n) \leq P_t, \forall n \in \mathcal{N},\\
& \mathbf{F} \succeq \mathbf{0},\label{P4_c_3}
\end{align}
\end{subequations}
where 
$\rho$ denotes the penalty factor.

However,
since the objective function (\ref{P4_obj}) is DC form,
we again adopt the SCA method to convexify the term $\Vert\mathbf{F}\Vert_{2}$
by linearization as follows
\begin{align}
\Vert\mathbf{F}\Vert_{2}
\geq
\Vert\mathbf{F}_0\Vert_{2}
+
\textrm{Tr}( \boldsymbol{\lambda}_{max}(\mathbf{F}_0)\boldsymbol{\lambda}_{max}^H(\mathbf{F}_0) (\mathbf{F} - \mathbf{F}_0)  ),
\end{align}
where
$\boldsymbol{\lambda}_{max}(\mathbf{F})$
denotes the eigenvector
corresponding to the largest eigenvalue of the matrix $\mathbf{F}$.
Therefore, 
the optimization problem (P4) can be further expressed  as follows
\begin{subequations}
\begin{align}
\textrm{(P5)}:&\mathop{\textrm{min}}
\limits_{\mathbf{F}, \{ t_g \}
}\
-{\sum}_{g=1}^{G} t_g + \frac{1}{2\rho} \Vert\mathbf{F}\Vert_{\ast} \label{P5_obj}\\
& - \frac{1}{2\rho}( \Vert\mathbf{F}_0\Vert_{2}
+
\textrm{Tr}( \boldsymbol{\lambda}_{max}(\mathbf{F}_0)\boldsymbol{\lambda}_{max}^H(\mathbf{F}_0) (\mathbf{F} - \mathbf{F}_0)  )  ) \nonumber
\\
\textrm{s.t.}\ 
&  \mathrm{\dot{R}}_k(\mathbf{F})   
- \textrm{Tr}( \nabla_{\mathbf{F}}^H\mathrm{\ddot{R}}_k(\mathbf{F}_0)(\mathbf{F}-\mathbf{F}_0) )\\
&- \mathrm{\ddot{R}}_k(\mathbf{F}_0)
 \geq t_g, 
\forall k \in \mathcal{K}_g, 
\forall g \in \mathcal{G}, \nonumber\\
& \text{Tr}(\mathbf{F}\mathbf{\bar{A}}_n) \leq P_t, \forall n \in \mathcal{N},\\
& \mathbf{F} \succeq \mathbf{0}.
\end{align}
\end{subequations}

The problem (P5) is convex w.r.t. the optimization variable $\mathbf{F}$, 
and hence it can be solved
by existing convex optimization solvers,
e.g., CVX \cite{ref_CVX}.

The proposed beamforming design algorithm can be summarized in Algorithm \ref{alg:1}.

\begin{algorithm}[t]
\caption{The Penalty-based Method}
\label{alg:1}
\begin{algorithmic}[1]
\STATE {initialize}
$\mathbf{F}^{(0)}$
and
$t=0$
;
\REPEAT
\STATE update $\mathbf{F}^{(t+1)}$ by solving  (P5);
\STATE $t++$;
\UNTIL{$convergence$;}
\end{algorithmic}
\end{algorithm}

\section{SOCP-based Method}

According to \cite{ref_SDP_complexity},
since the SDP problem (P5) always incurs high computational complexity,
we propose an SOCP-based method that has a lower computational complexity to solve problem (P0).

\subsection{Problem Reformulation}

Firstly,
we adopt the WMMSE framework 
\cite{ref_WMMSE}
to convert the objective function (\ref{P0_obj}) of problem (P0).
Specifically,
by introducing auxiliary variables $\{\beta_k \}$ and $\{\omega_k \}$,
the function $\mathrm{R}_k(\mathbf{f}) $ can be  written into an equivalent
variation form presented in (\ref{WMMSE_transformation}).
\begin{figure*}
\begin{align}
&\mathrm{R}_k(\mathbf{f})
=  \textrm{log} \big(1+ \vert\mathbf{\bar{h}}_k^H  \mathbf{B}_g \mathbf{f} \vert^2 
[ {\sum}_{i\neq g}^{G}
(\mathbf{f}^H \mathbf{B}_i \mathbf{\bar{h}}_k\mathbf{\bar{h}}_k^H \mathbf{B}_i \mathbf{f} )
 + \sigma_k^2 ]^{-1}  \big) \label{WMMSE_transformation}\\
&=\mathop{\textrm{max}}
\limits_{
\omega_k\geq0
}
\bigg\{
\textrm{log}(\omega_k)-\omega_k\big( 
{\sum}_{i=1}^{G}
(\mathbf{f}^H \mathbf{B}_i \mathbf{\bar{h}}_k\mathbf{\bar{h}}_k^H \mathbf{B}_i \mathbf{f} )
 + \sigma_k^2
\big)^{-1}\mathbf{\bar{h}}_k^H \mathbf{B}_g \mathbf{f} + 1\bigg\} \nonumber \\
& = 
\mathop{\textrm{max}}
\limits_{
\omega_k\geq0,
\beta_k
}
\bigg\{
\underbrace{
\textrm{log}(\omega_k)-\omega_k\big( 
1-2\text{Re}\{ \beta_k^{\ast} \mathbf{\bar{h}}_k^H \mathbf{B}_g \mathbf{f}\}
+ \vert\beta_k\vert^2(
{\sum}_{i=1}^{G}
(\mathbf{f}^H \mathbf{B}_i \mathbf{\bar{h}}_k\mathbf{\bar{h}}_k^H \mathbf{B}_i \mathbf{f} )
 + \sigma_k^2)
\big) + 1}
\limits_{\mathrm{\tilde{R}}_k(\mathbf{f},\omega_k,\beta_k)} \bigg\} ,
\forall k \in \mathcal{K}_g, 
\forall g \in \mathcal{G}.\nonumber
\end{align}
\boldsymbol{\hrule}
\end{figure*}

Therefore,
the original problem (P0) can be equivalently converted to
\begin{subequations}
\begin{align}
\textrm{(P6)}:&\mathop{\textrm{max}}
\limits_{\mathbf{f},
\{\omega_k\},
\{\beta_k\}
}\
\bigg\{\! \mathrm{R}_{s} (\mathbf{f}) 
\!= \!\!
{\sum}_{g=1}^{G} \!
\mathop{\textrm{min}}
\limits_{ k \in \mathcal{K}_g } \!
\mathrm{\tilde{R}}_k(\mathbf{f},\omega_k,\beta_k)\!\!
 \bigg\}\label{P6_obj}\\
\textrm{s.t.}\ 
& \mathbf{f}^H\mathbf{\bar{A}}_n\mathbf{f} \leq P_t, \forall n \in \mathcal{N},\label{P6_c_1}
\end{align}
\end{subequations}

In the following, 
we adopt the block coordinate ascent (BCA) \cite{ref_BCA} method to solve the problem (P6).

\subsection{Optimizing auxiliary variables}

According to the derivation of WMMSE transformation,
when other variables are fixed,
the update of the auxiliary variables $\{\beta_k \}$ and $\{\omega_k \}$
have analytical solutions,
which are given as follows
\begin{align}
&\beta_k^{\star} = \frac{\mathbf{\bar{h}}_k^H \mathbf{B}_g \mathbf{f}}
{{\sum}_{i=1}^{G}
(\mathbf{f}^H \mathbf{B}_i \mathbf{\bar{h}}_k\mathbf{\bar{h}}_k^H \mathbf{B}_i \mathbf{f} )
 + \sigma_k^2},\label{beta_opt} \\
& \omega_k^{\star} = 1+ \frac{\mathbf{f}^H \mathbf{B}_g \mathbf{\bar{h}}_k \mathbf{\bar{h}}_k^H \mathbf{B}_g \mathbf{f}}
{{\sum}_{i\neq g}^{G}
(\mathbf{f}^H \mathbf{B}_i \mathbf{\bar{h}}_k\mathbf{\bar{h}}_k^H \mathbf{B}_i \mathbf{f} )
 + \sigma_k^2}.\label{omega_opt}
\end{align}

\subsection{Updating The Beamformer}
In this subsection,
we discuss the update of the transmit beamformer $\mathbf{f}$
when other variables are given. 
By introducing the new coefficients as follows
\begin{align}
& \mathbf{b}_{1,k}
\! \triangleq \!
\omega_k\beta_k \mathbf{B}_g\mathbf{\bar{h}}_k,
\mathbf{B}_{1,k}
\!\triangleq\!\!
{\sum}_{i=1}^{G}
\omega_k\vert \beta_k \vert^2 ( \mathbf{B}_i \mathbf{\bar{h}}_k\mathbf{\bar{h}}_k^H \mathbf{B}_i  ),\\
&c_{1,k}
\!\triangleq\!
\text{log}(\omega_k)\! -\! \omega_k
\!+ \!2\omega_k\text{Re}\{ \beta_k^{\ast} \mathbf{\bar{h}}_k^H \mathbf{B}_g \mathbf{f} \}
\!-\! \omega_k\vert \beta_k \vert^2  \sigma_k^2  + 1,\nonumber
\end{align}
the function $\mathrm{\tilde{R}}_k(\mathbf{f},\omega_k,\beta_k)$ is equivalently rewritten as
\begin{align}
\mathrm{\tilde{R}}_k
=
-\mathbf{f}^H\mathbf{B}_{1,k}\mathbf{f} + 2\text{Re}\{ \mathbf{b}_{1,k}^H \mathbf{f} \} + c_{1,k}. \label{P7_obj_rewritten}
\end{align}

Based on the above transformation, 
the beamformer optimization problem can be expressed as
\begin{subequations}
\begin{align}
\textrm{(P7)}:&\mathop{\textrm{max}}
\limits_{\mathbf{f}
}\ \!\!\!
{\sum}_{g=1}^{G} 
\mathop{\textrm{min}}
\limits_{ k \in \mathcal{K}_g } 
\!
\{\!-\!\mathbf{f}^H\!\mathbf{B}_{1,k}\mathbf{f}\! +\! 2\text{Re}\{ \mathbf{b}_{1,k}^H \mathbf{f} \}\! +\! c_{1,k}\!\}
\label{P7_obj}\\
\textrm{s.t.}\
& \mathbf{f}^H\mathbf{\bar{A}}_n\mathbf{f} \leq P_t, \forall n \in \mathcal{N}.\label{P7_c_1}
\end{align}
\end{subequations}

Furthermore,
by introducing the slack variables $\{ \hat{t}_g \}$,
the optimization problem (P7) can be given as
\begin{subequations}
\begin{align}
\textrm{(P8)}:&\mathop{\textrm{max}}
\limits_{\mathbf{f},
\{\omega_k\},
\{\beta_k\},
\{ \hat{t}_g \}
}\
{\sum}_{g=1}^{G}\hat{t}_g \label{P8_obj}\\
\textrm{s.t.}\
&
\mathbf{f}^H\mathbf{B}_{1,k}\mathbf{f} - 2\text{Re}\{ \mathbf{b}_{1,k}^H \mathbf{f} \} - c_{1,k} + \hat{t}_g \leq 0, \\
& \mathbf{f}^H\mathbf{\bar{A}}_n\mathbf{f} \leq P_t, \forall n \in \mathcal{N},\label{P8_c_2}
\end{align}
\end{subequations}

The problem (P8) is a typical SOCP
and can be solved by  CVX.
The SOCP-based method can be summarized in Algorithm \ref{alg:2}.

\begin{algorithm}[t]
\caption{The SOCP-based Method}
\label{alg:2}
\begin{algorithmic}[1]
\STATE {initialize}
$\mathbf{f}^{(0)}$
and
$t=0$
;
\REPEAT
\STATE update $\{\beta_k^{(t+1)}\}$ and $\{\omega_k^{(t+1)}\}$ by (\ref{beta_opt}) and (\ref{omega_opt}), respectively;
\STATE update $\mathbf{f}^{(t+1)}$ by solving  (P8);
\STATE $t++$;
\UNTIL{$convergence$;}
\end{algorithmic}
\end{algorithm}

\section{Low-complexity Algorithm}

Note that our previously proposed Alg.1 and Alg.2 significantly rely on numerical solvers, 
e.g., CVX, 
to update the transmit beamformer. 
This reliance may lead to some undesirable properties:

i) General convex optimization solvers, including CVX,
relies on interior point (IP) method \cite{ref_Convex Optimization} to resolve SDP and/or SOCP
problems, 
which always produces high computational complexity.

ii) The use of third-party solvers inherently raises costs and adds inconvenience to algorithm implementation, 
such as the need for purchasing licenses, 
installing and maintaining software, 
and ensuring the necessary platform support for the solver.

Therefore, we proceed to explore an efficient solution that, hopefully,
does not rely on any numerical solvers.

\subsection{Efficient Update of $\mathbf{f}$}
In this subsection,
we investigate a low-complexity solution for solving the problem (P7) in Sec. IV-C. 
First, note that the coefficient $\mathbf{B}_{1,k}$ is a block diagonal matrix. 
Combining this with the structure of the TRTC's element power constraint (\ref{P7_c_1}),
we decompose the variable $\mathbf{f}$ into $N$ subvariables $\{\mathbf{\bar{f}}_{n}\}$,
each of which is defined as follows
\begin{align}
\mathbf{\bar{f}}_{n} \triangleq & [ \mathbf{f}( n), \mathbf{f}(N + n), \cdots, \mathbf{f}((g-1)\times N+ n) ,\\ 
&\cdots, \mathbf{f}((G-1)\times N+ n) ]^T\in \mathbb{C}^{G\times 1}. \nonumber
\end{align}

And then,
we introduce the following notations presented in (\ref{P9_notations}).
\begin{figure*}
\begin{align}
&\mathbf{\bar{b}}_{1,n,k} \triangleq [\mathbf{B}_{1,k}(n,n),\mathbf{B}_{1,k}(N+n,N+n), 
 \cdots,  \mathbf{B}_{1,k}((G-1)\times N+ n,(G-1)\times N+ n)  ]^T\in \mathbb{C}^{G\times 1},\label{P9_notations}\\
& \mathbf{\bar{B}}_{n,k} \triangleq \text{diag}(\mathbf{\bar{b}}_{1,n,k}),
\bar{b}_{2,n,k,g} \triangleq {\sum}_{j\neq n}^{N} \mathbf{f}^{\ast}((g-1)\times N+ j)\mathbf{B}_{1,k}((g-1)\times N+ j,(g-1)\times N+ n),\nonumber\\
&\mathbf{b}_{2,n,k}\triangleq [\bar{b}_{2,n,k,1}, \bar{b}_{2,n,k,2},\cdots,\bar{b}_{2,n,k,G}]^T\in \mathbb{C}^{G\times 1},
\mathbf{b}_{3,n,k}\triangleq[ \mathbf{b}_{1,k}( n ),\mathbf{b}_{1,k}( N + n ),\cdots,\mathbf{b}_{1,k}( (G-1)\times N+ n )  ]^T\in \mathbb{C}^{G\times 1},\nonumber\\
&c_{2,n,k}\triangleq {\sum}_{g=1}^{G}
{\sum}_{i\neq n}^{N}
{\sum}_{j\neq n}^{N}
 \mathbf{f}^{\ast}((g-1)\times N+ i) 
\mathbf{B}_{1,k}((g-1)\times N+ i,(g-1)\times N+ j) 
\mathbf{f}((g-1)\times N+ j),\nonumber\\
&c_{3,n,k}\triangleq {\sum}_{i \neq n}^{N} 2\text{Re}\{  \mathbf{b}_{3,i,k}^H \mathbf{\bar{f}}_{i} \}, 
\mathbf{b}_{4,n,k} \triangleq \mathbf{b}_{3,n,k} - \mathbf{b}_{2,n,k},
c_{4,n,k} \triangleq c_{1,k} - c_{2,n,k} + c_{3,n,k}.\nonumber
\end{align}
\boldsymbol{\hrule}
\end{figure*}
With other variables (i.e., $\{\mathbf{\bar{f}}_{i}, i\neq n \}$) being fixed,
the function $\mathrm{\tilde{R}}_k(\mathbf{f},\omega_k,\beta_k)$ can be rewritten as
\begin{align}
\mathrm{\tilde{R}}_k
=
\underbrace{
-\mathbf{\bar{f}}_{n}^H\mathbf{\bar{B}}_{n,k}\mathbf{\bar{f}}_{n} + 2\text{Re}\{ \mathbf{b}_{4,n,k}^H\mathbf{\bar{f}}_{n} \} + c_{4,n,k}
}\limits_{ \mathrm{\acute{R}}_{n,k} (\mathbf{\bar{f}}_{n}) },
\end{align}

Therefore,
the optimization problem w.r.t. the variable $\mathbf{\bar{f}}_{n}$ can be  formulated as
\begin{subequations}
\begin{align}
\textrm{(P9)}:&\mathop{\textrm{max}}
\limits_{\mathbf{\bar{f}}_{n}
}\ 
{\sum}_{g=1}^{G} 
\mathop{\textrm{min}}
\limits_{ k \in \mathcal{K}_g } 
\{\mathrm{\acute{R}}_{n,k} (\mathbf{\bar{f}}_{n})\}
\label{P9_obj}\\
\textrm{s.t.}\
& \mathbf{\bar{f}}_{n}^H\mathbf{\bar{f}}_{n} \leq P_t.\label{P9_c_1}
\end{align}
\end{subequations}

Obviously,
 the objective function $ \mathop{\textrm{min}}_{ k \in \mathcal{K}_g } 
\{\mathrm{\acute{R}}_{n,k} (\mathbf{\bar{f}}_{n})\}$ in (\ref{P9_obj})  is non-differentiable.
Based on the smooth approximation theory \cite{ref_Log_sum},
we can approximate it as follows
\begin{align}
&\mathop{\textrm{min}}_{ k \in \mathcal{K}_g } 
\{\mathrm{\acute{R}}_{n,k} (\mathbf{\bar{f}}_{n})\}
\approx
\mathrm{\breve{R}}_{n,g} (\mathbf{\bar{f}}_{n})\\
&=
-\frac{1}{\mu_{n,g}} \text{log}\bigg( \sum_{k \in \mathcal{K}_g} \text{exp} \big( -\mu_{n,g} \mathrm{\acute{R}}_{n,k} (\mathbf{\bar{f}}_{n})  
\big) \bigg),\nonumber
\end{align}
where
$\mathrm{\breve{R}}_{n,g} (\mathbf{\bar{f}}_{n})$
is a smooth function and is the lower bound for 
 $ \mathop{\textrm{min}}_{ k \in \mathcal{K}_g } 
\mathrm{\acute{R}}_{n,k} (\mathbf{\bar{f}}_{n})$,
and
$\mu_{n,g}$ is a smoothing parameter that satisfies the following condition:
\begin{align}
\mathrm{\breve{R}}_{n,g}(\mathbf{\bar{f}}_{n})\!\! +\!\! \frac{1}{\mu_{n,g}} \text{log}(\vert \mathcal{K}_g \vert)
\!\!\geq \!\!
\mathop{\textrm{min}}_{ k \in \mathcal{K}_g } 
\{\mathrm{\acute{R}}_{n,k}(\mathbf{\bar{f}}_{n})\}
\!\!\geq\!\!
\mathrm{\breve{R}}_{n,g}(\mathbf{\bar{f}}_{n}).
\end{align}

Besides,
the function $-\frac{1}{\mu_{n,g}} \text{log}\big( \sum_{k \in \mathcal{K}_g} \text{exp} \big( -\mu_{n,g} \mathrm{\acute{R}}_{n,k} (\mathbf{\bar{f}}_{n})  
\big) \big)$ is monotonically increasing and concave w.r.t. 
$\mathrm{\acute{R}}_{n,k} (\mathbf{\bar{f}}_{n})$, which has been proven in \cite{ref_RIS_A_2}.
And the function $\mathrm{\acute{R}}_{n,k} (\mathbf{\bar{f}}_{n})$ 
is concave in $\mathbf{\bar{f}}_{n}$.
Based on the composition principle \cite{ref_Convex Optimization},
$\mathrm{\breve{R}}_{n,g}(\mathbf{\bar{f}}_{n})$ is also a concave function of $\mathbf{\bar{f}}_{n}$.

When the appropriate value of $\mu_{n,g}$ is given,
we turn to solve the following problem
\begin{subequations}
\begin{align}
\textrm{(P10)}:&\mathop{\textrm{max}}
\limits_{\mathbf{\bar{f}}_{n}
}\ 
{\sum}_{g=1}^{G} 
\mathrm{\breve{R}}_{n,g}(\mathbf{\bar{f}}_{n})
\label{P10_obj}\\
\textrm{s.t.}\
& \mathbf{\bar{f}}_{n}^H\mathbf{\bar{f}}_{n} \leq P_t.\label{P10_c_1}
\end{align}
\end{subequations}

It can be observed that 
the above problem (P10) is still complex and difficult to solve.
Inspired by the MM framework \cite{ref_MM}, 
we can construct a lower bound of the objective function (\ref{P10_obj}), 
which is given as
\begin{align}
&\mathrm{\breve{R}}_{n,g}(\mathbf{\bar{f}}_{n})
\geq 
\mathrm{\grave{R}}_{n,g}(\mathbf{\bar{f}}_{n}\vert \mathbf{\bar{f}}_{n,0})\label{MM_lower_bound}\\
&= c_{5,n,g} + 2\text{Re}\{\mathbf{b}_{5,n,g}^H \mathbf{\bar{f}}_{n}\} + \alpha_{n,g}\mathbf{\bar{f}}_{n}^H\mathbf{\bar{f}}_{n}, \nonumber 
\end{align}
where
$\mathbf{\bar{f}}_{n,0}$ is obtained from the last iteration,
and the newly introduced coefficients are defined as follows
\begin{align}
& h_{n,k}( \mathbf{\bar{f}}_{n,0} ) \triangleq 
\frac{ \text{exp}( - \mu_{n,g} \mathrm{\acute{R}}_{n,k} (\mathbf{\bar{f}}_{n})  )  } 
{\sum_{k\in \mathcal{K}_g}  \text{exp} ( - \mu_{n,g} \mathrm{\acute{R}}_{n,k} (\mathbf{\bar{f}}_{n})  )  }, \label{MM_coefficient} \\
& tp_{n,k} 
\triangleq 
\lambda_{\text{max}}(\mathbf{\bar{B}}_{n,k}\mathbf{\bar{B}}_{n,k}^H)P_{t}
\!+\! \Vert \mathbf{b}_{4,n,k} \Vert_2^2
\!+\! 2\sqrt{P_{t}} \Vert \mathbf{\bar{B}}_{n,k}\mathbf{b}_{4,n,k}  \Vert_2
,\nonumber\\
&\alpha_{n,g}
\triangleq 
- \mathop{\textrm{max}}
\limits_{ k \in \mathcal{K}_g } \{\lambda_{\text{max}}(\mathbf{\bar{B}}_{n,k}) \}
-2\mu_{n,g}\mathop{\textrm{max}}
\limits_{ k \in \mathcal{K}_g }
\{ tp_{n,k}  \}
,\nonumber\\
& \mathbf{b}_{5,n,g}
\triangleq 
{\sum}_{k\in \mathcal{K}_g}  h_{n,k}( \mathbf{\bar{f}}_{n,0} ) ( \mathbf{b}_{4,n,k} - \mathbf{\bar{B}}_{n,k}^H\mathbf{\bar{f}}_{n,0}  ) - \alpha_{n,g}\mathbf{\bar{f}}_{n,0},\nonumber\\ 
& c_{5,n,g} \triangleq \mathrm{\breve{R}}_{n,g}(\mathbf{\bar{f}}_{n,0}) 
- 2\text{Re}\{ \mathbf{b}_{6,n,g}^H  \mathbf{\bar{f}}_{n,0} \}  + \alpha_{n,g}\mathbf{\bar{f}}_{n,0}^H\mathbf{\bar{f}}_{n,0},\nonumber
\end{align}
and the derivation details of (\ref{MM_coefficient}) can be seen in Appendix A.

Therefore, 
replacing the function $\mathrm{\breve{R}}_{n,g}(\mathbf{\bar{f}}_{n})$ by (\ref{MM_lower_bound}),
we turn to optimize a convex lower bound of the objective function  of (P10),
which is expressed as
\begin{subequations}
\begin{align}
\textrm{(P11)}:&\mathop{\textrm{max}}
\limits_{\mathbf{\bar{f}}_{n}
}\ 
\bar{\alpha}_{n}\mathbf{\bar{f}}_{n}^H\mathbf{\bar{f}}_{n} + 2\text{Re}\{\mathbf{b}_{7,n}^H \mathbf{\bar{f}}_{n}\}  + c_{6,n} 
\label{P11_obj}\\
&\textrm{s.t.}\
 \mathbf{\bar{f}}_{n}^H\mathbf{\bar{f}}_{n} \leq P_t.\label{P11_c_1}
\end{align}
\end{subequations}
where 
\begin{align}
&\bar{\alpha}_{n} \triangleq {\sum}_{g=1}^{G} \alpha_{n,g},
 \mathbf{b}_{7,n} \triangleq {\sum}_{g=1}^{G} \mathbf{b}_{5,n,g},\\
& c_{6,n} \triangleq {\sum}_{g=1}^{G} c_{5,n,g}.\nonumber
\end{align}

Since $\bar{\alpha}_{n} \leq 0$,
the problem (P11) is convex 
and can be solved via off-the-shelf numerical solvers, e.g., CVX.

In order to efficiently solve the problem (P11), 
we adopt the Lagrangian multiplier method \cite{ref_Convex Optimization} to obtain the optimal closed-form solution of problem (P11).
Firstly,
by denoting the Lagrangian multiplier associated with the
constraint of (\ref{P11_c_1}) as $\nu$,
the Lagrange function associated with the
problem (P11) is written as
\begin{align}
\mathcal{L}(\mathbf{\bar{f}}_{n},\nu) =& 
 - \bar{\alpha}_{n}\mathbf{\bar{f}}_{n}^H\mathbf{\bar{f}}_{n} - 2\text{Re}\{\mathbf{b}_{6,n}^H \mathbf{\bar{f}}_{n}\}  \\
 &- c_{6,n} + \nu( \mathbf{\bar{f}}_{n}^H\mathbf{\bar{f}}_{n} - P_t ),\nonumber
\end{align}

Furthermore,
by setting the first-order derivative of the Lagrange function $\mathcal{L}(\mathbf{\bar{f}}_{n},\nu)$ w.r.t. $\mathbf{\bar{f}}_{n}$ to zero,
we can have 
\begin{align}
\frac{\partial \mathcal{L}(\mathbf{\bar{f}}_{n},\nu)}{\partial \mathbf{\bar{f}}_{n}} = \mathbf{0}.
\end{align}

And then,
we can obtain the solution of $\mathbf{\bar{f}}_{n}$ as follows
\begin{align}
\mathbf{\bar{f}}_{n} = \frac{\mathbf{b}_{6,n}}{ \nu - \bar{\alpha}_{n}  }. \label{P11_closed_solution}
\end{align}

By substituting the equation (\ref{P11_closed_solution}) into the power constraint (\ref{P11_c_1}), 
the resulting expression is formulated as follows
\begin{align}
 \frac{\mathbf{b}_{6,n}^H\mathbf{b}_{6,n}}{ (\nu - \bar{\alpha}_{n})^2  }\leq P_t. \label{MM_closed_power}
\end{align}
Note that 
the left hand side of (\ref{MM_closed_power}) is a decreasing function w.r.t. 
the Lagrangian multiplier $\nu$.
Then
the optimal solution to problem (P11) is given by one of the
following two cases:
\begin{itemize}
\item[] \underline{CASE-I}:
If the equation (\ref{MM_closed_power}) is satisfied when $\nu = 0$,
then the optimal solution of (P11) is given by
\begin{align}
\mathbf{\bar{f}}_{n}^{\star} = -\frac{\mathbf{b}_{6,n}}{  \bar{\alpha}_{n}  }. 
\end{align}

\item[] \underline{CASE-II}: 
Otherwise,
$\nu $ is positive.
And the optimal solution to problem (P11) becomes
\begin{align}
\mathbf{\bar{f}}_{n}^{\star} = \sqrt{P_t}\frac{\mathbf{b}_{6,n}}{  \Vert\mathbf{b}_{6,n}\Vert_2  }. 
\end{align}

\end{itemize}

The MM-based method can be summarized in Algorithm \ref{alg:3},
where
$ \mathcal{R} (\cdot) $
is the objective function (\ref{P0_obj})
and
$ \mathcal{F} (\cdot) $
denotes the nonlinear fixed-point iteration map of
the MM-based method in (\ref{P11_closed_solution}).

\begin{algorithm}[t]
\caption{The MM-based Method}
\label{alg:3}
\begin{algorithmic}[1]
\STATE {initialize}
$\mathbf{f}^{(0)}$
and
$t=0$
;
\REPEAT
\STATE update $\{\beta_k^{(t+1)}\}$ and $\{\omega_k^{(t+1)}\}$ by (\ref{beta_opt}) and (\ref{omega_opt}), respectively;
\FOR{$n = 1$ to $N$}
\STATE $\mathbf{\bar{f}}_{n,1} = \mathcal{F} (\mathbf{\bar{f}}_{n}^{(t)} )$;
\STATE $\mathbf{\bar{f}}_{n,2} = \mathcal{F} (\mathbf{\bar{f}}_{n,1} )$;
\STATE $\mathbf{j}_{1} = \mathbf{\bar{f}}_{n,1} - \mathbf{\bar{f}}_{n}^{(t)}$;
\STATE $\mathbf{j}_{2} = \mathbf{\bar{f}}_{n,2} - \mathbf{\bar{f}}_{n,1} - \mathbf{j}_{1}$;
\STATE $\tau = - \frac{\Vert \mathbf{j}_{1}\Vert_2}{\Vert\mathbf{j}_{2}\Vert_2} $;
\STATE $\mathbf{\bar{f}}_{n}^{(t+1)} = \mathbf{\bar{f}}_{n}^{(t)} - 2\tau\mathbf{j}_{1} + \tau^2\mathbf{j}_{2} $;
\STATE if $\Vert\mathbf{\bar{f}}_{n}^{(t+1)}\Vert_2^2 > P_t$,
  $ \mathbf{\bar{f}}_{n}^{(t+1)} = \sqrt{P_t}  \frac{\mathbf{\bar{f}}_{n}^{(t+1)}}{\Vert\mathbf{\bar{f}}_{n}^{(t+1)}\Vert_2}  ; 
  $
\WHILE{ $\mathcal{R} (\mathbf{\bar{f}}_{n}^{(t+1)}) < \mathcal{R} (\mathbf{\bar{f}}_{n}^{(t)})$ }
\STATE $\tau = (\tau -1)/2$;
\STATE if $\Vert\mathbf{\bar{f}}_{n}^{(t+1)}\Vert_2^2 > P_t$,
  $ \mathbf{\bar{f}}_{n}^{(t+1)} = \sqrt{P_t} \frac{\mathbf{\bar{f}}_{n}^{(t+1)}}{\Vert\mathbf{\bar{f}}_{n}^{(t+1)}\Vert_2}   $;
\ENDWHILE
\ENDFOR
\STATE $t++$;
\UNTIL{$convergence$;}
\end{algorithmic}
\end{algorithm}

\section{Numerical Results}

\begin{figure}[t]
	\centering
	\includegraphics[width=.35\textwidth]{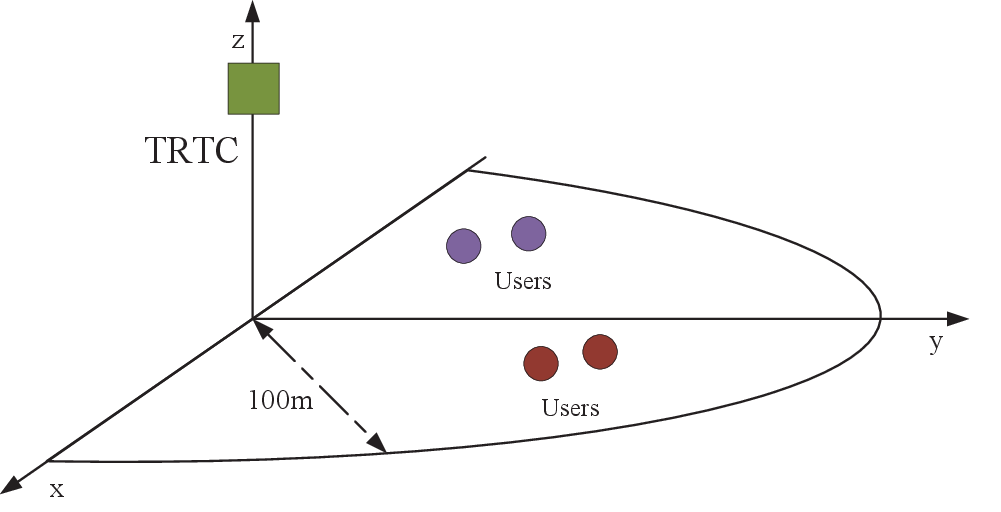}
	\caption{Simulation setup for a multigroup multicast MISO
communication system using a TRTC.}
	\label{fig.2}
\end{figure}

In this section, 
extensive simulation results are provided to validate 
the effectiveness of the proposed algorithms for 
the considered TRTC-enabled downlink multigroup multicast MISO communication system.
The  setting of the simulation is shown in  Fig. \ref{alg:2}.
It contains one TRTC and $K = 4$ mobile users, evenly divided into $G = 2$ groups.
In the experiment, 
the TRTC is located at the three-dimensional (3D) coordinates (0,0,4.5).
All users are randomly distributed within a right half circle of radius 100m centered at the TRTC, 
and are placed at a height of 1.5m.
The antenna spacing is set to half the wavelength of the carrier.
The large-scale fading is given as 
\begin{align}
\text{PL} = C_0 \bigg (\frac{d}{d_0} \bigg)^{-\alpha},  
\end{align}
where $C_0$ denotes the path loss of the reference distance $d_0 = 1$m, 
and $d$ and $\alpha $ represent the propagation distance and the fading exponent, respectively.
The TRTC-user link
adopts the Rician distribution with a Rician factor of 5dB.
The path loss exponent of the TRTC-user link is $\alpha_{l} = 3.6$.
The transmit power for each unit of the TRTC is set as 10dBm. 
The noise power  is  set as $\sigma^2 = -90$dBm.
The threshold for algorithm convergence is set as $\varepsilon = 10^{-4}$.

\begin{figure}[t]
	\centering
	\includegraphics[width=.5\textwidth]{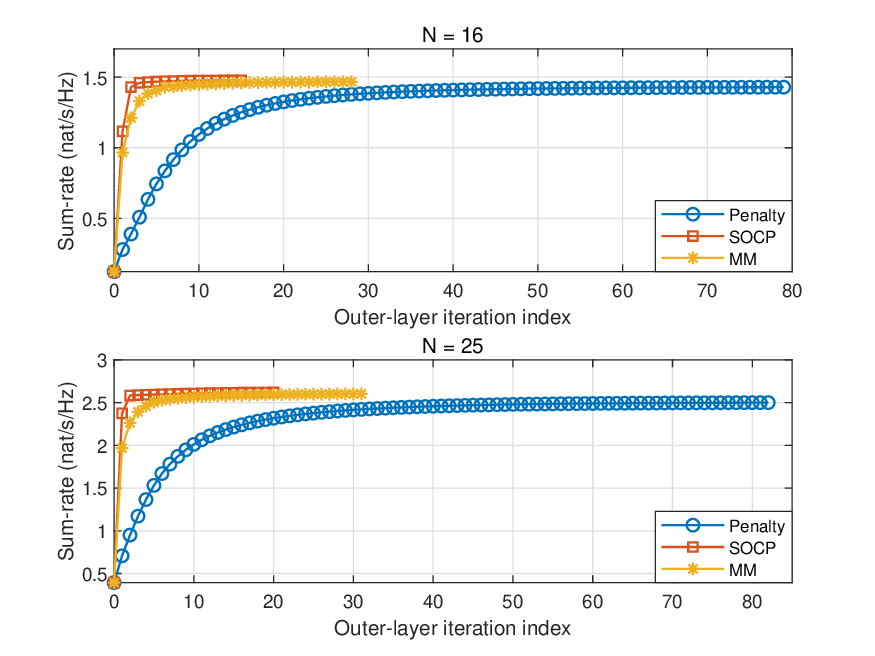}
	\caption{Convergence of Algorithms.}
	\label{fig.3}
\end{figure}

First, 
we label the proposed  algorithm \ref{alg:1}, \ref{alg:2} and \ref{alg:3} as
``Penalty", ``SOCP" and ``MM", respectively.
For fair comparison,
three algorithm implementations start from one common initial point in each channel realization.
Fig. \ref{fig.3} presents the convergence behavior of our proposed algorithms.
The upper and
lower subplots correspond to different unit numbers for TRTC, respectively. 
It is observed that the sum-rate achieved by three solutions 
monotonically increases with the iteration index, 
exhibiting notably rapid improvement during the initial iterations.
After convergence is achieved, 
the SOCP-based solution achieves the highest sum-rate performance, 
while the penalty-based algorithm yields the lowest. 
The MM-based algorithm exhibits only a slight performance degradation compared to the SOCP solution.
As seen from the figure, both SOCP-based and MM-based solutions  generally converge within 10 iterations,
and the penalty-based algorithm can converge within 50 iterations.

\begin{figure}[t]
	\centering
	\includegraphics[width=.5\textwidth]{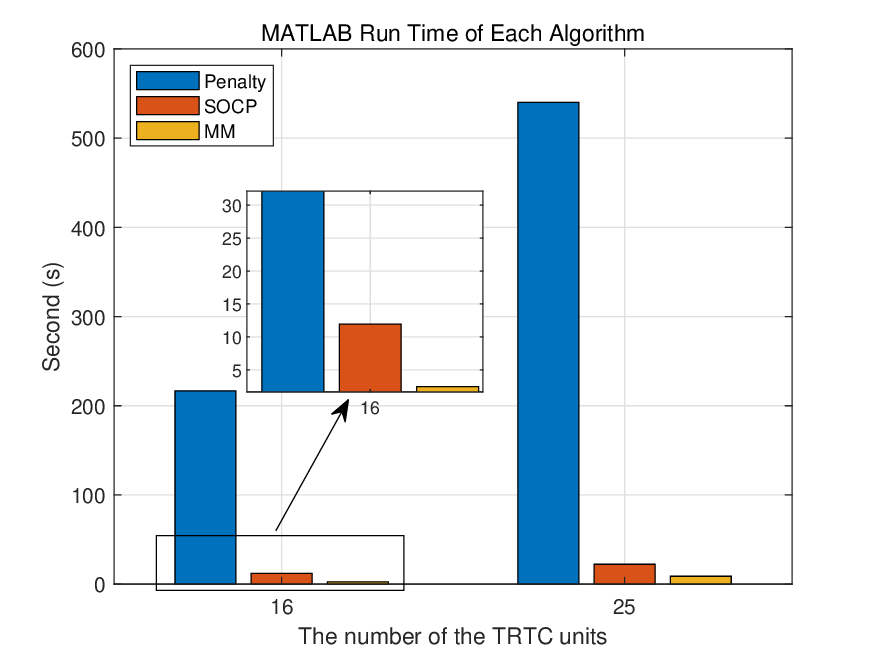}
	\caption{Comparison of MATLAB Run Time for Three Algorithms.}
	\label{fig.4}
\end{figure}

Furthermore, 
we investigate the computational complexity of our proposed three algorithms in the above convergence experiment. 
Under different settings of the TRTC element numbers $N$,
the MATLAB runtime comparisons for three algorithms are presented in Fig. \ref{fig.4}.
As shown by the results,
the runtime of the penalty-based method is the longest, 
followed by that of the SOCP-based algorithm, 
and the MM-based algorithm's runtime is the shortest.
The runtime of the ``SOCP'' method is generally two orders of magnitude lower than that of the penalty-based algorithm, 
while the ``MM'' algorithm demonstrates a further reduction in runtime 
by approximately one order of magnitude compared to the ``SOCP'' method.
As shown in Fig.\ref{fig.3} and Fig.\ref{fig.4}, 
although the MM-based method converges to a slightly lower sum-rate compared to the SOCP-based method, 
it achieves a significantly shorter runtime.

\begin{figure}[t]
	\centering
	\includegraphics[width=.5\textwidth]{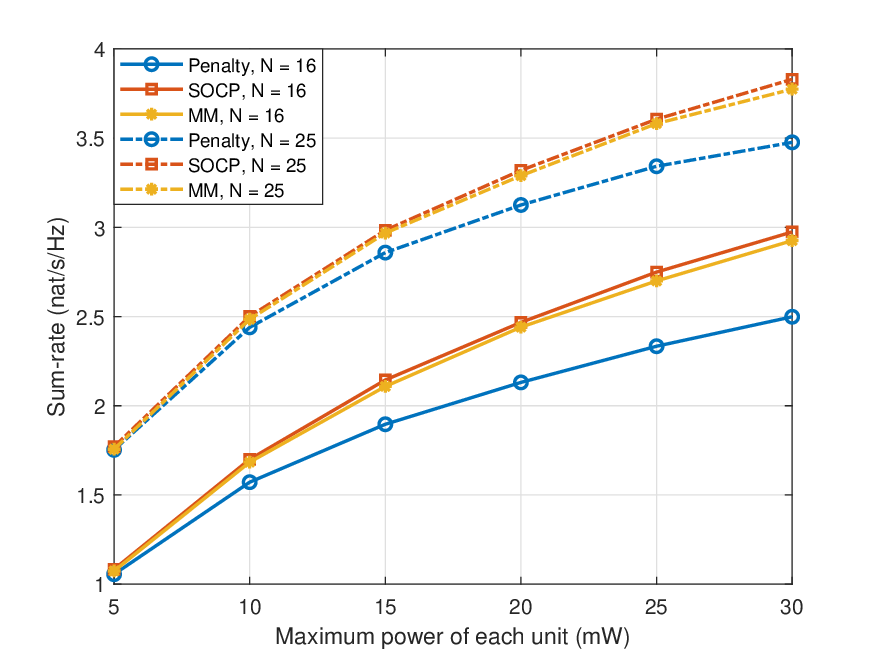}
	\caption{Sum-rate versus the maximum power of
each transmissive unit.}
	\label{fig.5}
\end{figure}

In Fig. \ref{fig.5}, 
we illustrate the sum-rate performance of three proposed algorithms versus the maximum transmit power of each TRTC unit.
It is clearly observed that, 
as the maximum transmit power of the TRTC unit gradually increases, 
the sum-rate increases monotonically for all three proposed schemes, 
demonstrating the effectiveness of power enhancement.
Both the SOCP and/or MM-based methods significantly outperform the penalty-based method.
Furthermore,
the gap between the SOCP and/or MM-based methods and the penalty-based method 
gradually becomes larger as the maximum transmit power of the TRTC unit continues to increase over a wide range of values.
Moreover, 
the sum-rate in the ``N = 25'' case significantly outperforms that in the ``N = 16'' case 
across all three considered algorithms under the same conditions.

\begin{figure}[t]
	\centering
	\includegraphics[width=.5\textwidth]{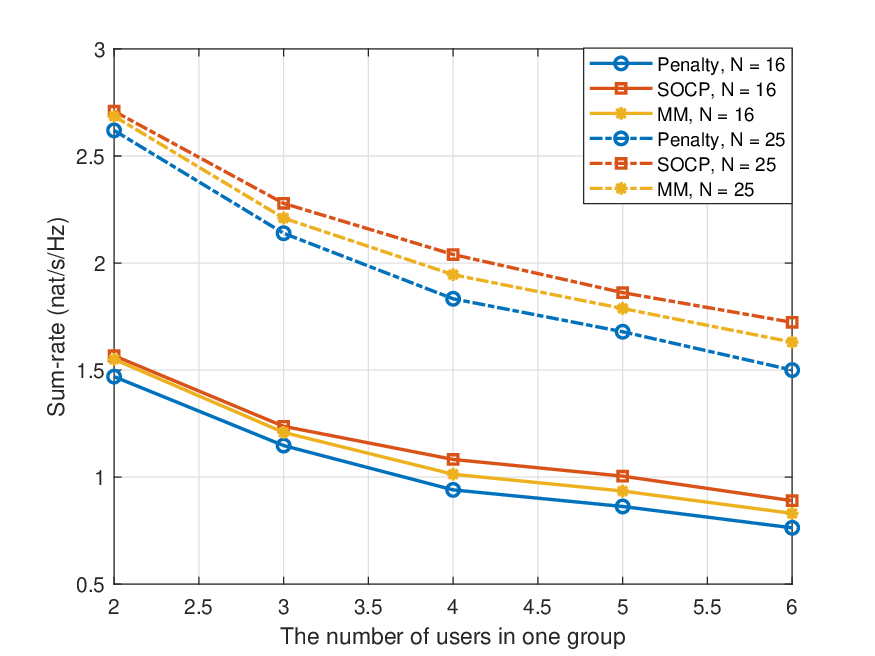}
	\caption{Sum-rate versus the user number of
each group.}
	\label{fig.6}
\end{figure}

Fig. \ref{fig.6} demonstrates the achievable sum-rate performance of three methods versus the number of users in each group.
Interestingly,
it is observed that 
the sum-rate of all cases decreases as the number of user in each group increases.
Among the three proposed algorithms, the SOCP-based method achieves the best performance, followed by the MM-based method, 
while the penalty-based method exhibits the lowest sum-rate performance.
As the number of users within each group increases, 
the gap in sum-rate performance between the SOCP-based and MM-based methods becomes increasingly pronounced, 
and the performance gap between the MM-based and penalty-based methods also gradually widens.
Moreover, 
under the identical user number, 
all three proposed algorithms achieve significantly higher sum-rates in the case of ``N = 25'' compared to the case of ``N = 16''.

\begin{figure}[t]
	\centering
	\includegraphics[width=.5\textwidth]{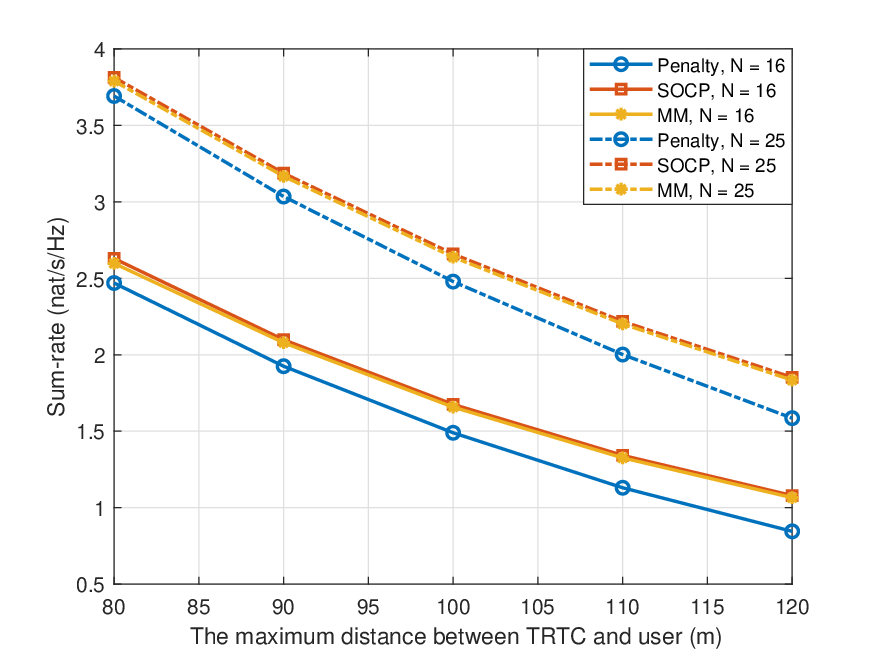}
	\caption{Sum-rate versus the maximum distance between TRTC and user.}
	\label{fig.7}
\end{figure}

The effect of the maximum distance between TRTC and user on the performance of all algorithms is shown in Fig. \ref{fig.7}.
Initially, 
it is observed that as the maximum distance between TRTC and user increases from $80$m to $120$m, 
there is a consistent decrease in the sum-rate across all schemes.
Moreover, 
the gap between the SOCP and/or MM-based and penalty-based methods gradually increases as the maximum distance between TRTC and user increases.
Given the same system setting, 
the sum-rate performance of all three proposed algorithms is considerably higher when N = 25 than when N = 16, 
and the gap between the two cases gradually decreases as the maximum distance between TRTC and user increases.

\begin{figure}[t]
	\centering
	\includegraphics[width=.5\textwidth]{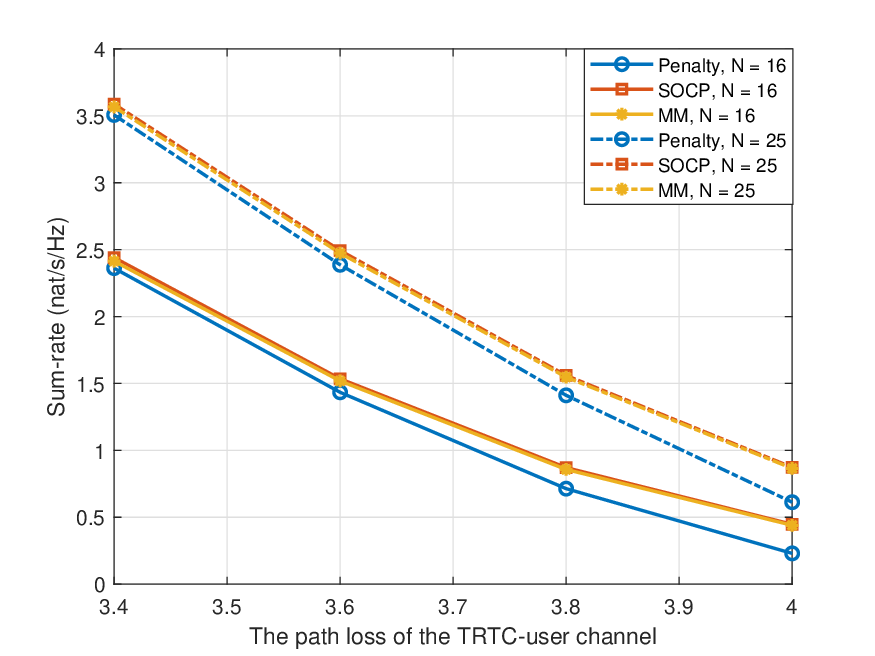}
	\caption{Sum-rate versus the path loss exponent.}
	\label{fig.8}
\end{figure}

Fig. \ref{fig.8} investigates the impact of the path loss exponent of the TRTC-user channel on the sum-rate.
As the path loss exponent $\alpha_{l}$ increases from $3.4$ to $4.0$,
the sum rate achieved by all considered schemes exhibits a clear and consistent monotonic decline.
In addition,
it is also observed that the performance gap between the SOCP/MM-based methods and the penalty-based methods gradually narrows as the path loss exponent $\alpha_{l}$ increases.
Furthermore, 
the sum-rate performance of all schemes is significantly improved when the number of TRTC elements is increased from $16$ to $25$.
However, 
it is also noteworthy that the sum rate gap between ``N = 16'' and ``N = 25'' cases also decreases as $\alpha_{l}$ increases.

\begin{figure}[t]
	\centering
	\includegraphics[width=.5\textwidth]{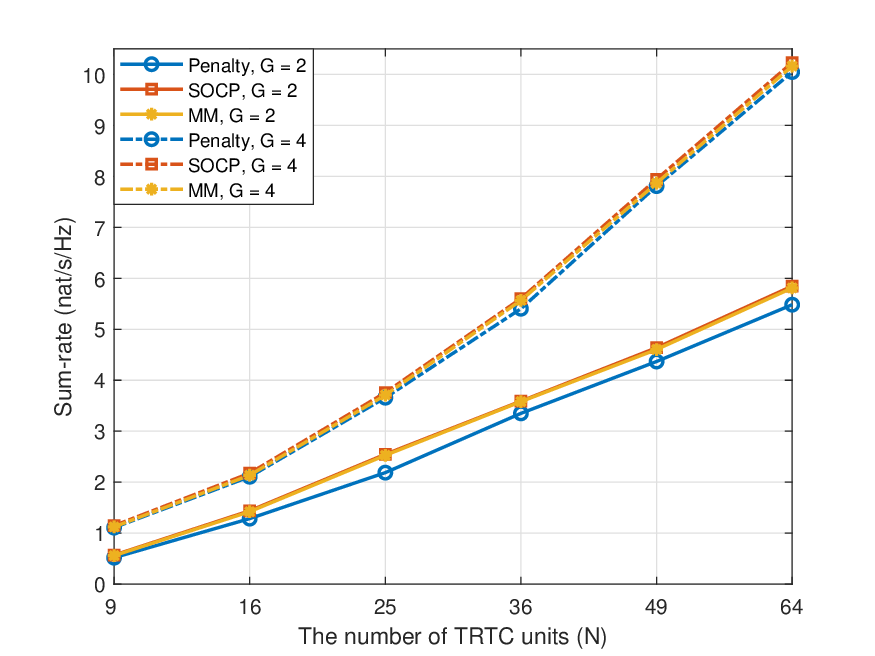}
	\caption{Sum-rate versus the number of TRTC elements.}
	\label{fig.9}
\end{figure}

Fig. \ref{fig.9} illustrates the impact of the number of TRTC elements.
Clearly, 
increasing the number of units can improve beamforming gain for all algorithms.
Additionally,
the sum-rate's 
growth rate w.r.t. $N$ in the case of $G = 2$ is much lower than that for $G = 4$.

\section{Conclusions}

This paper investigates a TRTC-enabled multigroup multicast MISO communication system, 
aiming to maximize the minimum rate among all user groups 
by optimizing the transmit beamforming vectors at the TRTC, 
subject to individual transmit power constraints at each TRTC unit.
To solve the max-min rate optimization problem while balancing performance and computational complexity, 
we propose three iterative algorithms: a penalty-based algorithm, an SOCP-based algorithm, and an MM-based algorithm.
Numerical results demonstrate that all three proposed optimization methods significantly enhance the sum-rate performance 
and highlight the potential of TRTC as a novel transceiver architecture for wireless systems characterized by low cost and low power consumption.
Furthermore, it is shown that the MM-based approach offers an efficient solution for transmit beamforming optimization, 
achieving reduced complexity with only a slight performance trade-off.

\appendix
\subsection{Proof of  (\ref{MM_lower_bound})}
\normalem
Proof:
First,
we briefly introduce the MM framework \cite{ref_MM}.
The MM method aims to simplify the complex optimization problem by constructing a surrogate objective function and/or the constraints.
These surrogates are easier to optimize and are used in place of the original objective function and/or constraints at each iteration.
 Specifically, 
 let $f(\mathbf{x})$ denote the original objective function, 
 and let $\mathbb{S}_{\mathbf{x}}$ denote the feasible set, which is assumed to be convex w.r.t. the variable $\mathbf{x}$.
Let $\mathbf{x}^{t-1}$  denotes
the optimal solution at the $t-1$th iteration,
and let $u(\mathbf{x}  \vert\mathbf{x}^{t-1})$ represents
a function of variable $ \mathbf{x} $ with given $\mathbf{x}^{t-1}$.
The convex approximation $u(\mathbf{x}  \vert\mathbf{x}^{t})$  
should satisfy the following conditions:
\begin{align}
&C1): u(\mathbf{x}^{t}  \vert\mathbf{x}^{t}) = f(\mathbf{x}^{t}), \forall \mathbf{x}^{t} \in  \mathbb{S}_{\mathbf{x}}; \\
&C2): f(\mathbf{x}) \geq u(\mathbf{x}  \vert\mathbf{x}^{t}), \forall \mathbf{x}^{t}, \mathbf{x} \in  \mathbb{S}_{\mathbf{x}};\nonumber\\
&C3): \nabla_{\mathbf{x}^{t}}u(\mathbf{x}^{t}  \vert\mathbf{x}^{t})  =  \nabla_{\mathbf{x}^{t}} f(\mathbf{x}^{t});\nonumber\\
&C4): u(\mathbf{x}  \vert\mathbf{x}^{t})\ \text{is continuous in}\ \mathbf{x}\ \text{and}\ \mathbf{x}^{t}.\nonumber
\end{align}

The first condition represents that the convex approximation function 
$u(\mathbf{x}^{t}  \vert\mathbf{x}^{t})$
and the original function $f(\mathbf{x}^{t})$ should be equal.
 The second condition states that
the original function serves as an upper bound of the convex approximation function. 
The third condition requires that 
the first-order gradient of the convex approximation function should be the same as that of the original function.

Note that the function $\mathrm{\breve{R}}_{n,g} (\mathbf{\bar{f}}_{n})$ is twice differentiable and concave.
Next,
via resorting to the MM method,
we consider a quadratic surrogate function to minorize the function $\mathrm{\breve{R}}_{n,g} (\mathbf{\bar{f}}_{n})$,
which is given as
\begin{align}
&\mathrm{\breve{R}}_{n,g} (\mathbf{\bar{f}}_{n})\geq
\mathrm{ \grave{R} }_{n,g}\label{Appendix_A_surrogate function} \\
&\triangleq
\mathrm{\breve{R}}_{n,g} (\mathbf{\bar{f}}_{n,0})
+ 2\text{Re}\{ \mathbf{b}_{6,n,g}^H ( \mathbf{\bar{f}}_{n} - \mathbf{\bar{f}}_{n,0} ) \}\nonumber \\
&+ ( \mathbf{\bar{f}}_{n} - \mathbf{\bar{f}}_{n,0} )^H \mathbf{M}_{n,g} ( \mathbf{\bar{f}}_{n} - \mathbf{\bar{f}}_{n,0} ),\nonumber
\end{align}
where
$\mathbf{b}_{6,n,g} \in \mathbb{C}^{G \times 1} $
and 
$\mathbf{M}_{n,g} \in \mathbb{C}^{G \times G}$.
Note that the function $\mathrm{ \hat{R} }_{n,g}$ should satisfy the MM method's conditions $C1)-C4)$.

Clearly,
both conditions $C1)$ and $C4)$ are already satisfied.
Next, 
we will sequentially verify that conditions $C3)$ and $C2)$ also hold.
Denote $\mathbf{\mathbf{ \tilde{f} }_{n}}$ belongs to $\mathbb{S}_{\mathbf{f}}$.
With given the direction $\mathbf{\mathbf{ \tilde{f} }_{n}} - \mathbf{\bar{f}}_{n,0}  $,
the directional derivative of the function $\mathrm{ \breve{R} }_{n,g}$ at the point $\mathbf{\bar{f}}_{n,0}$ 
can be written as
\begin{align}
2\text{Re}\big\{
\big( {\sum}_{k \in \mathcal{K}_g } h_{n,k}(\mathbf{\bar{f}}_{n,0} )
( \mathbf{b}_{4,n,k}^H - \mathbf{\bar{f}}_{n,0}^H \mathbf{\bar{B}}_{n,k} )   
\big)
(\mathbf{\mathbf{ \tilde{f} }_{n}} - \mathbf{\bar{f}}_{n,0})
\big\}.\label{Appendix_A_C2_1}
\end{align}

The  directional derivative of the function $\mathrm{ \grave{R} }_{n,g}$ is given by
\begin{align}
2\text{Re}\{ \mathbf{b}_{6,n,g}^H (\mathbf{\mathbf{ \tilde{f} }_{n}} - \mathbf{\bar{f}}_{n,0}) \}.\label{Appendix_A_C2_2}
\end{align}

To fulfill the condition $C3)$, 
the two directional derivatives given in  (\ref{Appendix_A_C2_1}) and (\ref{Appendix_A_C2_2}) must be equal, 
implying that
\begin{align}
\mathbf{b}_{6,n,g}
=
{\sum}_{k \in \mathcal{K}_g } h_{n,k}(\mathbf{\bar{f}}_{n,0} )
( \mathbf{b}_{4,n,k} -  \mathbf{\bar{B}}_{n,k}^H\mathbf{\bar{f}}_{n,0} ).
\end{align}

We now proceed to verify that the condition $C2)$ is also satisfied. 
When
 the surrogate function $\mathrm{ \grave{R} }_{n,g}(\mathbf{\bar{f}}_{n} \vert\mathbf{\bar{f}}_{n,0} )$
serves as a lower bound for every linear segment in any direction, 
the condition $C2)$ holds.
Let 
$\mathbf{\bar{f}}_{n} = \mathbf{\bar{f}}_{n,0} + \gamma( \mathbf{ \tilde{f} }_{n} - \mathbf{\bar{f}}_{n,0} )$, 
$\forall \gamma \in [0,1]$.
And then, 
the following expression needs to be fulfilled
\begin{align}
&\mathrm{\breve{R}}_{n,g} (\mathbf{\bar{f}}_{n,0} + \gamma( \mathbf{ \tilde{f} }_{n} - \mathbf{\bar{f}}_{n,0} ))\label{Appendix_A_C3_1} \\
&\geq
\mathrm{\breve{R}}_{n,g} (\mathbf{\bar{f}}_{n,0} ) 
+ 2\gamma\text{Re}\{ \mathbf{b}_{6,n,g}^H ( \mathbf{\bar{f}}_{n} - \mathbf{\bar{f}}_{n,0} ) \} \nonumber \\
&+ \gamma^2( \mathbf{\bar{f}}_{n} - \mathbf{\bar{f}}_{n,0} )^H \mathbf{M}_{n,g} ( \mathbf{\bar{f}}_{n} - \mathbf{\bar{f}}_{n,0} ).\nonumber
\end{align}

We define the functions
$L_{n,g}(\gamma) \triangleq \mathrm{\breve{R}}_{n,g} (\mathbf{\bar{f}}_{n,0} + \gamma( \mathbf{ \tilde{f} }_{n}
 - \mathbf{\bar{f}}_{n,0} ))$
and
$l_{n,k}(\gamma) \triangleq  \mathrm{\acute{R}}_{n,k} (\mathbf{\bar{f}}_{n,0} + \gamma( \mathbf{ \tilde{f} }_{n}
 - \mathbf{\bar{f}}_{n,0} ))   $.
A sufficient condition for (\ref{Appendix_A_C3_1}) is given as
\begin{align}
\frac{ \partial^2 L_{n,g}(\gamma)  }{\partial  \gamma^2} \geq 
2( \mathbf{\bar{f}}_{n} - \mathbf{\bar{f}}_{n,0} )^H \mathbf{M}_{n,g} ( \mathbf{\bar{f}}_{n} - \mathbf{\bar{f}}_{n,0} ).
\label{Appendix_A_C3_1_1}
\end{align}

First,
the first-order derivative of $L_{n,g}(\gamma)$ is expressed as
\begin{align}
\frac{ \partial L_{n,g}(\gamma)  }{\partial  \gamma} 
=
{\sum}_{k \in \mathcal{K}_g}h_{1,n,k}(\gamma)\nabla_{\gamma} l_{n,k}(\gamma),\label{Appendix_A_C3_3}
\end{align}
where
\begin{align}
&h_{1,n,k}(\gamma) \triangleq \frac{\text{exp}( -\mu_{n,g}l_{n,k}(\gamma) )}
{ {\sum}_{k\in \mathcal{K}_g}  \text{exp}( -\mu_{n,g}l_{n,k}(\gamma) ) },\\
&\nabla_{\gamma} l_{n,k}(\gamma)\triangleq
-2\gamma ( \mathbf{ \tilde{f} }_{n} - \mathbf{\bar{f}}_{n,0} )^H
\mathbf{\bar{B} }_{n,k}
( \mathbf{ \tilde{f} }_{n} - \mathbf{\bar{f}}_{n,0} )\nonumber\\
&+ 2\text{Re}\{ \mathbf{b}_{4,n,k}^H( \mathbf{ \tilde{f} }_{n} - \mathbf{\bar{f}}_{n,0} ) 
- \mathbf{\bar{f}}_{n,0}^H  \mathbf{\bar{B} }_{n,k}
( \mathbf{ \tilde{f} }_{n} - \mathbf{\bar{f}}_{n,0} ) \}
\nonumber\\
& = 2\text{Re}\{ \mathbf{e}_{n,k}^H \mathbf{ \hat{f} }_{n}\},
\mathbf{ \hat{f} }_{n}
\triangleq
\mathbf{ \tilde{f} }_{n} - \mathbf{\bar{f}}_{n,0}
\nonumber\\
& 
\mathbf{e}_{n,k} 
\triangleq
\mathbf{b}_{4,n,k}-\mathbf{\bar{B} }_{n,k}^H(\mathbf{\bar{f}}_{n,0} + \gamma( \mathbf{ \tilde{f} }_{n} - \mathbf{\bar{f}}_{n,0} )  ).\nonumber
\end{align}

Furthermore,
the second-order derivative is given in (\ref{Appendix_A_C3_2}),
\begin{figure*}
\begin{align}
\frac{ \partial^2 L_{n,g}(\gamma)  }{\partial  \gamma^2} 
=
{\sum}_{k\in \mathcal{K}_g}\big(  h_{1,n,k}(\gamma)  \nabla_{\gamma}^2 l_{n,k}(\gamma) 
- \mu_{n,g}h_{1,n,k}(\gamma)  (\nabla_{\gamma} l_{n,k}(\gamma) )^2
 \big)+  \mu_{n,g}\big( {\sum}_{k\in \mathcal{K}_g} h_{1,n,k}(\gamma)\nabla_{\gamma} l_{n,k}(\gamma) \big)^2.\label{Appendix_A_C3_2}
\end{align}
\boldsymbol{\hrule}
\end{figure*}
where 
\begin{align}
\nabla_{\gamma}^2 l_{n,k}(\gamma) 
&=
-2( \mathbf{ \tilde{f} }_{n} - \mathbf{\bar{f}}_{n,0} )^H
\mathbf{\bar{B} }_{n,k}
( \mathbf{ \tilde{f} }_{n} - \mathbf{\bar{f}}_{n,0} )\label{Appendix_A_C3_4}\\
&=-2  \mathbf{ \hat{f} }_{n}^H \mathbf{\bar{B} }_{n,k}\mathbf{ \hat{f} }_{n},\nonumber
\end{align}

Based on the equations (\ref{Appendix_A_C3_3})$-$(\ref{Appendix_A_C3_4}),
the second-order derivative $\frac{ \partial^2 L_{n,g}(\gamma)  }{\partial  \gamma^2} $
can be rewritten as
\begin{align}
\frac{ \partial^2 L_{n,g}(\gamma)  }{\partial  \gamma^2}
=
\begin{bmatrix}
\mathbf{ \hat{f} }_{n}^H & \mathbf{ \hat{f} }_{n}^T
\end{bmatrix}
\boldsymbol{\Phi}_{n,g}
\begin{bmatrix}
\mathbf{ \hat{f} }_{n} \\
\mathbf{ \hat{f} }_{n}^{\ast}
\end{bmatrix},
\end{align}
with
the newly introduced coefficient 
$\boldsymbol{\Phi}_{n,g}$ defined in
(\ref{Appendix_A_C3_5}).
\begin{figure*}
\begin{align}
\boldsymbol{\Phi}_{n,g}
&\triangleq
\sum_{k \in \mathcal{K}_g}
\bigg(
h_{1,n,k}(\gamma) 
\begin{bmatrix}
- \mathbf{ \bar{B}}_{n,k} & \mathbf{0} \\ 
        \mathbf{0}        & - \mathbf{ \bar{B}}_{n,k}
\end{bmatrix}
-
\mu_{n,g}h_{1,n,k}(\gamma) 
\begin{bmatrix}
 \mathbf{ e}_{n,k}\\ 
 \mathbf{ e}_{n,k}^{\ast}
\end{bmatrix}
\begin{bmatrix}
 \mathbf{ e}_{n,k}\\ 
 \mathbf{ e}_{n,k}^{\ast}
\end{bmatrix}^H
\bigg)
\label{Appendix_A_C3_5}\\
&+\mu_{n,g}
\begin{bmatrix}
\sum_{k \in \mathcal{K}_g} h_{1,n,k}(\gamma)   \mathbf{ e}_{n,k}\\ 
\sum_{k \in \mathcal{K}_g} h_{1,n,k}(\gamma)  \mathbf{ e}_{n,k}^{\ast}
\end{bmatrix}
\begin{bmatrix}
\sum_{k \in \mathcal{K}_g} h_{1,n,k}(\gamma)   \mathbf{ e}_{n,k}\\ 
\sum_{k \in \mathcal{K}_g} h_{1,n,k}(\gamma)  \mathbf{ e}_{n,k}^{\ast}
\end{bmatrix}^H. \nonumber 
\end{align}
\boldsymbol{\hrule}
\end{figure*}

Similarly,
the right  of the inequality (\ref{Appendix_A_C3_1_1}) is reexpressed as
\begin{align}
&2( \mathbf{\bar{f}}_{n} - \mathbf{\bar{f}}_{n,0} )^H \mathbf{M}_{n,g} ( \mathbf{\bar{f}}_{n} - \mathbf{\bar{f}}_{n,0} )\\
&=
\begin{bmatrix}
\mathbf{ \hat{f} }_{n}^H & \mathbf{ \hat{f} }_{n}^T
\end{bmatrix}
\begin{bmatrix}
 \mathbf{M}_{n,g} & \mathbf{0} \\
\mathbf{0}  &  \mathbf{M}_{n,g}
\end{bmatrix}
\begin{bmatrix}
\mathbf{ \hat{f} }_{n} \\
\mathbf{ \hat{f} }_{n}^{\ast}
\end{bmatrix}.\nonumber
\end{align}

To fulfill
the condition $C2)$,
we have
\begin{align}
\boldsymbol{\Phi}_{n,g}
\succeq
\begin{bmatrix}
 \mathbf{M}_{n,g} & \mathbf{0} \\
\mathbf{0}  &  \mathbf{M}_{n,g}
\end{bmatrix}.
\end{align}

When we choose $\mathbf{M}_{n,g} = \alpha_{n,g} \mathbf{I} = \lambda_{\text{min}}(\boldsymbol{\Phi}_{n,g})\mathbf{I} $,
the function $\mathrm{ \grave{R} }_{n,g}$ in (\ref{Appendix_A_surrogate function})
can be formulated as
\begin{align}
&\mathrm{ \grave{R} }_{n,g}
=
\mathrm{\breve{R}}_{n,g} (\mathbf{\bar{f}}_{n,0})
+ 2\text{Re}\{ \mathbf{b}_{6,n,g}^H ( \mathbf{\bar{f}}_{n} - \mathbf{\bar{f}}_{n,0} ) \} \\
&+ ( \mathbf{\bar{f}}_{n} - \mathbf{\bar{f}}_{n,0} )^H \mathbf{M}_{n,g} ( \mathbf{\bar{f}}_{n} - \mathbf{\bar{f}}_{n,0} )\nonumber\\
&= c_{5,n,g} + 2\text{Re}\{\mathbf{b}_{5,n,g}^H \mathbf{\bar{f}}_{n}\} + \alpha_{n,g}\mathbf{\bar{f}}_{n}^H\mathbf{\bar{f}}_{n},\nonumber
\end{align}
where
$c_{5,n,g}$
and
$\mathbf{b}_{5,n,g}$
are defined in (\ref{MM_coefficient}).

Since the matrix $\boldsymbol{\Phi}_{n,g}$ is complex,
the value of $\alpha_{n,g}$ is difficult to obtain.
Next,
we introduce the following lemmas for obtaining the value of $\alpha_{n,g}$,
which are formulated as

\begin{itemize}
\item[] {a1)}:
Given that the matrices $\mathbf{A}$ and $\mathbf{B}$ are Hermitian,
the inequality $ \lambda_{\text{min}}(\mathbf{A}) + \lambda_{\text{min}}(\mathbf{B}) \leq  \lambda_{\text{min}}(\mathbf{A}+\mathbf{B}) $
holds;

\item[] {a2)}: 
If the matrix $\mathbf{A}$ has rank one,
$ \lambda_{\text{max}}(\mathbf{A}) = \text{Tr}(\mathbf{A})$
and
$ \lambda_{\text{min}}(\mathbf{A}) = 0$;

\item[] {a3)}: 
When
$a_k, b_k \geq 0$
and
$\sum_{k=1}^{K} a_k = 1$,
we have
$ \sum_{k=1}^{K}a_kb_k \leq \text{max}_{k=1}^{K}b_k $;

\item[] {a4)}: 
Let $\mathbf{A}$ and $\mathbf{B}$ be positive semidefinite matrices, 
with $\mathbf{A}$ having maximum eigenvalue $\lambda_{\text{max}}(\mathbf{A})$. 
Then the following inequality holds:
$ \text{Tr}(\mathbf{A}\mathbf{B}) \leq \lambda_{\text{max}}(\mathbf{A})\text{Tr}(\mathbf{B}) $.
\end{itemize}

By leveraging $a1)  - {a4)}$,
we can obtain a lower bound of $\alpha_{n,g}$,
which the derivation procedure is formulated in (\ref{Appendix_A_C3_6}).
\begin{figure*}
\begin{align}
\lambda_{\text{min}}(\boldsymbol{\Phi}_{n,g})
&
\overset{ \text{a1)} }
\geq
\sum_{k \in \mathcal{K}_g}
h_{1,n,k}(\gamma) 
\lambda_{\text{max}}
\bigg(
\begin{bmatrix}
- \mathbf{ \bar{B}}_{n,k} & \mathbf{0}\label{Appendix_A_C3_6} \\ 
        \mathbf{0}        & - \mathbf{ \bar{B}}_{n,k}
\end{bmatrix}
\bigg)
-
\sum_{k \in \mathcal{K}_g}
\mu_{n,g}h_{1,n,k}(\gamma) 
\lambda_{\text{max}}
\bigg(
\begin{bmatrix}
 \mathbf{ e}_{n,k}\\ 
 \mathbf{ e}_{n,k}^{\ast}
\end{bmatrix}
\begin{bmatrix}
 \mathbf{ e}_{n,k}\\ 
 \mathbf{ e}_{n,k}^{\ast}
\end{bmatrix}^H
\bigg)
\\
&+
\mu_{n,g}
\lambda_{\text{min}}
\bigg(
\begin{bmatrix}
\sum_{k \in \mathcal{K}_g} h_{1,n,k}(\gamma)   \mathbf{ e}_{n,k}\\ 
\sum_{k \in \mathcal{K}_g} h_{1,n,k}(\gamma)  \mathbf{ e}_{n,k}^{\ast}
\end{bmatrix}
\begin{bmatrix}
\sum_{k \in \mathcal{K}_g} h_{1,n,k}(\gamma)   \mathbf{ e}_{n,k}\\ 
\sum_{k \in \mathcal{K}_g} h_{1,n,k}(\gamma)  \mathbf{ e}_{n,k}^{\ast}
\end{bmatrix}^H
\bigg)\nonumber\\
&
\overset{ \text{a2)} }
=
-
\sum_{k \in \mathcal{K}_g}
h_{1,n,k}(\gamma) 
\big(
\lambda_{\text{max}}(\mathbf{ \bar{B}}_{n,k})
+
2\mu_{n,g}
\mathbf{ e}_{n,k}^H\mathbf{ e}_{n,k}
\big)
\overset{ \text{a3)} }
\geq
-
\underset{k \in \mathcal{K}_g}
{\text{max}}
\{ \lambda_{\text{max}}(\mathbf{ \bar{B}}_{n,k}) \}
-
2\mu_{n,g}
\underset{k \in \mathcal{K}_g}
{\text{max}}
\{ 
\Vert
\mathbf{e}_{n,k}
\Vert_2^2
\}.
\nonumber
\end{align}
\boldsymbol{\hrule}
\end{figure*}

Note that the value of $\Vert
\mathbf{e}_{n,k}
\Vert_2^2$ in (\ref{Appendix_A_C3_6})
is still difficult to obtain.
Next,
we proceed to find its upper bound.
Since
$\mathbf{\bar{f}}_{n} = \mathbf{\bar{f}}_{n,0} + \gamma( \mathbf{ \tilde{f} }_{n} - \mathbf{\bar{f}}_{n,0} )$, 
$\forall \gamma \in [0,1]$,
the inequality 
$\Vert \mathbf{\bar{f}}_{n}\Vert_2^2 = 
\Vert \mathbf{\bar{f}}_{n,0} + \gamma( \mathbf{ \tilde{f} }_{n} - \mathbf{\bar{f}}_{n,0} ) \Vert_2^2
\leq P_t$
holds.
By leveraging 
$a4)$, 
an upper bound of 
the term 
$\Vert
\mathbf{e}_{n,k}
\Vert_2^2$
can be seen in (\ref{Appendix_A_C3_7}).
\begin{figure*}
\begin{align}
&\Vert
\mathbf{e}_{n,k}
\Vert_2^2
=
\Vert
\mathbf{b}_{4,n,k}-\mathbf{\bar{B} }_{n,k}^H(\mathbf{\bar{f}}_{n,0} + \gamma( \mathbf{ \tilde{f} }_{n} - \mathbf{\bar{f}}_{n,0} )  )
\Vert_2^2\label{Appendix_A_C3_7}\\
&=
\Vert
\mathbf{b}_{4,n,k}
\Vert_2^2
+
\Vert
\mathbf{\bar{B} }_{n,k}^H(\mathbf{\bar{f}}_{n,0} + \gamma( \mathbf{ \tilde{f} }_{n} - \mathbf{\bar{f}}_{n,0} )  )
\Vert_2^2
-
2\text{Re}\{
\mathbf{b}_{4,n,k}^H
\mathbf{\bar{B} }_{n,k}^H(\mathbf{\bar{f}}_{n,0} + \gamma( \mathbf{ \tilde{f} }_{n} - \mathbf{\bar{f}}_{n,0} )  )
\}\nonumber\\
&
\overset{a4)}
\leq
\lambda_{\text{max}}(\mathbf{\bar{B} }_{n,k}\mathbf{\bar{B} }_{n,k}^H)
\Vert
\mathbf{\bar{f}}_{n,0} + \gamma( \mathbf{ \tilde{f} }_{n} - \mathbf{\bar{f}}_{n,0} )  
\Vert_2^2
+\Vert
\mathbf{b}_{4,n,k}
\Vert_2^2
-
2\text{Re}\{
\mathbf{b}_{4,n,k}^H
\mathbf{\bar{B} }_{n,k}^H(\mathbf{\bar{f}}_{n,0} + \gamma( \mathbf{ \tilde{f} }_{n} - \mathbf{\bar{f}}_{n,0} )  )
\}
\nonumber\\
&
\leq
\lambda_{\text{max}}(\mathbf{\bar{B} }_{n,k}\mathbf{\bar{B} }_{n,k}^H)
P_t
+\Vert
\mathbf{b}_{4,n,k}
\Vert_2^2
+
2\sqrt{P_t}\Vert\mathbf{\bar{B} }_{n,k}\mathbf{b}_{4,n,k}\Vert_2.\nonumber
\end{align}
\boldsymbol{\hrule}
\end{figure*}
Specifically,
the last term $2\sqrt{P_t}\Vert\mathbf{\bar{B} }_{n,k}\mathbf{b}_{4,n,k}\Vert_2$ of
the last inequality in (\ref{Appendix_A_C3_7}) is the optimal solution of the following optimization problem,
which is given as
\begin{subequations}
\begin{align}
\mathop{\textrm{min}}
\limits_{\mathbf{x}
}\ 
& 2\text{Re}\{\mathbf{b}_{4,n,k}^H \mathbf{\bar{B} }_{n,k}^H \mathbf{x} \}
\\
\textrm{s.t.}\ &
 \mathbf{x}^H\mathbf{x} \leq P.
\end{align}
\end{subequations}

Finally, by combining (\ref{Appendix_A_C3_6}) and (\ref{Appendix_A_C3_7}), 
we can obtain the lower bound  of $\alpha_{n,g}$ in (\ref{MM_coefficient}).

Therefore, 
the coefficients in
(\ref{MM_coefficient}) have been proved.



\begin{thebibliography}{99}



\bibitem{ref_RIS_1}
C. Pan \emph{et al.}, 
``An overview of signal processing techniques for RIS/IRS-aided wireless systems,'' 
\emph{IEEE J. Sel. Topics Signal Process.}, 
vol. 16, no. 5, pp. 883$-$917, Aug. 2022.

\bibitem{ref_RIS_2}
Q. Wu, S. Zhang, B. Zheng, C. You, and R. Zhang, 
``Intelligent reflecting surface-aided wireless communications: A tutorial,'' 
\emph{IEEE Trans. Commun.}, 
vol. 69, no. 5, pp. 3313$-$3351, May 2021.

\bibitem{ref_RIS_3}
Q. Wu \emph{et al.}, 
``Intelligent surfaces empowered wireless network: Recent advances and the road to 6G,'' 
\emph{ Proc. IEEE}, 
vol. 112, no. 7, pp. 724$-$763, July 2024.

\bibitem{ref_RIS_A_1}
C. Pan \emph{ et al.}, 
``Multicell MIMO communications relying on intelligent reflecting surfaces,'' 
\emph{ IEEE Trans. Wireless Commun.}, 
vol. 19, no. 8, pp. 5218$-$5233, Aug. 2020.


\bibitem{ref_RIS_A_2}
G. Zhou, C. Pan, H. Ren, K. Wang, and A. Nallanathan, 
``Intelligent reflecting surface aided multigroup multicast MISO communication systems,'' 
\emph{ IEEE Trans. Signal Process.}, 
vol. 68, pp. 3236$-$3251, 2020.



\bibitem{ref_RIS_A_3}
Z. He, H. Shen, W. Xu, and C. Zhao, 
``Low-cost passive beamforming for RIS-aided wideband OFDM systems,''
\emph{ IEEE Wireless Commun. Lett.}, 
vol. 11, no. 2, pp. 318$-$322, Feb. 2022.



\bibitem{ref_RIS_A_4}
S. Gong, C. Xing, P. Yue, L. Zhao, and T. Q. S. Quek, 
``Hybrid analog and digital beamforming for RIS-assisted mmWave communications,'' 
\emph{ IEEE Trans. Wireless Commun.}, 
vol. 22, no. 3, pp. 1537$-$1554, Mar. 2023.



\bibitem{ref_RIS_A_5}
Z. Peng, R. Weng, C. Pan, G. Zhou, M. D. Renzo, and A. L. Swindlehurst, 
``Robust transmission design for RIS-assisted secure multiuser communication systems in the presence of hardware impairments,'' 
\emph{ IEEE Trans. Wireless Commun.}, 
vol. 22, no. 11, pp. 7506$-$7521, Nov. 2023.



\bibitem{ref_RIS_A_6}
Y. Guo, Y. Liu, M. Li, Q. Wu, and Q. Shi, 
``Beamforming design for power transferring and secure communication in RIS-aided network,'' 
in \emph{Proc. IEEE Int. Conf. Commun. (ICC)}, 
Seoul, Korea, Republic of, 2022, pp. 450$-$455. 



\bibitem{ref_RIS_A_7}
S. Li, H. Du, D. Zhang, and K. Li, 
``Joint UAV trajectory and beamforming designs for RIS-assisted MIMO system,'' 
\emph{ IEEE Trans. Veh. Technol.},
vol. 73, no. 4, pp. 5378$-$5392, April 2024.



\bibitem{ref_RIS_A_8}
J. Wang, J. Xiao, Y. Zou, W. Xie, and Y. Liu, 
``Wideband beamforming for RIS assisted near-field communications,'' 
\emph{ IEEE Trans. Wireless Commun.}, 
vol. 23, no. 11, pp. 16836$-$16851, Nov. 2024.



\bibitem{ref_RIS_A_9}
Z. Wang, X. Hu, C. Liu, and M. Peng, 
``RIS-enabled multi-target sensing: Performance analysis and space-time beamforming design,'' 
\emph{ IEEE Trans. Wireless Commun.}, 
vol. 23, no. 10, pp. 13889$-$13903, Oct. 2024.



\bibitem{ref_RIS_A_10}
Y. Guo, Y. Liu, Q. Wu, X. Li, and Q. Shi, 
``Joint beamforming and power allocation for RIS aided full-duplex integrated sensing and uplink communication system,'' 
\emph{ IEEE Trans. Wireless Commun.}, 
vol. 23, no. 5, pp. 4627$-$4642, May 2024.


\bibitem{ref_RIS_A_11}
Z. Zhang, W. Chen, Q. Wu, Z. Li, X. Zhu, and J. Yuan, 
``Intelligent omni surfaces assisted integrated multi-target sensing and multi-user MIMO communications,'' 
\emph{ IEEE Trans. Commun.}, 
vol. 72, no. 8, pp. 4591$-$4606, Aug. 2024.

\bibitem{ref_RIS_A_12}
Z. Liu, Y. Liu, S. Shen, Q. Wu, and Q. Shi, 
``Enhancing ISAC network throughput using beyond diagonal RIS,'' 
\emph{IEEE Wireless Commun. Lett.}, 
vol. 13, no. 6, pp. 1670$-$1674, June 2024.

\bibitem{ref_RIS_A_13}
Z. Guang, Y. Liu, Q. Wu, W. Wang, and Q. Shi, 
``Power minimization for ISAC system using beyond diagonal reconfigurable intelligent surface,'' 
\emph{ IEEE Trans. Veh. Technol.},
vol. 73, no. 9, pp. 13950$-$13955, Sept. 2024.


\bibitem{ref_RIS_A_14}
X. Yang, Z. Wei, Y. Liu, H. Wu, and Z. Feng, 
``RIS-assisted cooperative multicell ISAC systems: A multi-user and multi-target case,'' 
\emph{ IEEE Trans. Wireless Commun.}, 
vol. 23, no. 8, pp. 8683$-$8699, Aug. 2024.






\bibitem{ref_TRIS_1}
Z. Li \emph{ et al.}, 
``Transmissive reconfigurable intelligent surface-enabled transceiver systems: Architecture, design issues, and opportunities,'' 
\emph{ IEEE Veh. Technol. Mag.}, 
vol. 19, no. 4, pp. 44$-$53, Dec. 2024.


\bibitem{ref_RRIS_1}
X. Bai, F. Kong, Y. Sun, G. Wang, J. Qian, X. Li, A. Cao, C. He, X. Liang, R. Jin, and W. Zhu, 
``High-efficiency transmissive programmable metasurface for multimode OAM generation,'' 
\emph{Adv. Opt. Mater.}, 
vol. 8, no. 17, p. 2000570, Jun. 2020.



\bibitem{ref_RRIS_2}
X. Bai, F. Zhang, L. Sun, A. Cao, J. Zhang, C. He, L. Liu, J. Yao, and W. Zhu, 
``Time-modulated transmissive programmable metasurface for low sidelobe beam scanning,''  
\emph{Research}, 
Jul. 2022.



\bibitem{ref_TRIS_A_1}
Z. Li \emph{ et al.}, 
``Toward TMA-based transmissive RIS transceiver enabled downlink communication networks: A consensus-ADMM approach,'' 
\emph{ IEEE Trans. Commun.}, 
vol. 73, no. 4, pp. 2832$-$2846, April 2025.



\bibitem{ref_TRIS_A_2}
Z. Li \emph{ et al.}, 
``Toward transmissive RIS transceiver enabled uplink communication systems: Design and optimization,''
\emph{IEEE Internet Things J.},
vol. 11, no. 4, pp. 6788$-$6801, Feb. 2024.



\bibitem{ref_TRIS_A_3}
Z. Li, W. Chen, Z. Zhang, Q. Wu, H. Cao, and J. Li, 
``Robust sum-rate maximization in transmissive RMS transceiver-enabled SWIPT networks,'' 
\emph{IEEE Internet Things J.},
vol. 10, no. 8, pp. 7259$-$7271,  April 2023.



\bibitem{ref_TRIS_A_4}
Z. Li, W. Chen, Z. Liu, H. Tang, and J. Lu, 
``Joint communication and computation design in transmissive RMS transceiver enabled multi-tier computing networks,'' 
\emph{IEEE J. Sel. Areas  Commun.}, 
vol. 41, no. 2, pp. 334$-$348, Feb. 2023.



\bibitem{ref_TRIS_A_5}
A. Huang, X. Mu, L. Guo, and G. Zhu, 
``Hybrid active-passive RIS transmitter enabled energy-efficient multi-user communications,'' 
\emph{ IEEE Trans. Wireless Commun.}, 
vol. 23, no. 9, pp. 10653$-$10666, Sept. 2024.



\bibitem{ref_TRIS_A_6}
R. Xiong, K. Yin, J. Lu, K. Wan, T. Mi, and R. C. Qiu,
``Flexible multi-beam synthesis and directional suppression through transmissive RIS,'' 
Nov. 2024.
[Online]. 
Available: https://arxiv.org/abs/2411.02008

\bibitem{ref_TRIS_A_6_1}
Y. Wang, S. Yang, Z. Chu, B. Ji, M. Hua, and C. Li, 
``Robust weighted sum secrecy rate maximization for joint ITS- and IRS-assisted multi-antenna networks,'' 
\emph{IEEE Wireless Commun. Lett.}, 
vol. 14, no. 3, pp. 681$-$685, March 2025.






\bibitem{ref_TRIS_A_7}
Z. Liu \emph{ et al.}, 
``Enhancing robustness and security in ISAC network design: Leveraging transmissive reconfigurable intelligent surface with RSMA,'' 
\emph{IEEE Trans. Commun.}, 
early access,
March 31, 2025,
doi: 10.1109/TCOMM.2025.3555894.



\bibitem{ref_TRIS_A_8}
Z. Liu, W. Chen, Q. Wu, Z. Li, Q. Wu, N. Cheng, and J. Li,
``Beamforming design and multi-user scheduling in transmissive RIS enabled distributed cooperative ISAC networks with RSMA,''
Nov. 2024.
[Online]. 
Available: https://arxiv.org/abs/2411.10960

\bibitem{ref_TRIS_A_9}
M. Asif, X. Bao, Z. Ali, A. Ihsan, M. Ahmed, and X. Li, 
``Transmissive RIS-empowered LEO-satellite communications with hybrid-NOMA under residual hardware impairments,'' 
\emph{IEEE Trans. Green Commun. Netw.}, 
early access,
September 23, 2024,
doi: 10.1109/TGCN.2024.3466469.


\bibitem{ref_TRIS_A_10}
J. Liu \emph{et al.}, 
``TRIS-HAR: Transmissive reconfigurable intelligent surfaces-assisted human activity recognition using state space models,'' 
\emph{IEEE Internet Things J.},
early access,
May 28, 2025,
doi: 10.1109/JIOT.2025.3574568.






\bibitem{ref_Convex Optimization}
S. Boyd and L. Vandenberghe,
\emph{Convex Optimization.}
New York: Cambridge University Press, 2004.

\bibitem{ref_WMMSE}
Q. Shi, M. Razaviyayn, Z. -Q. Luo, and C. He,
``An iteratively weighted MMSE approach to distributed sum-utility maximization for a MIMO interfering broadcast channel,''
\emph{IEEE Trans. Signal Process.},
vol. 59, no. 9, pp. 4331$-$4340, Sept. 2011.

\bibitem{ref_Log_sum}
S. Xu, 
``Smoothing method for minimax problems,''
\emph{Comput. Optim. Appl.}, 
vol. 20, no. 3, pp. 267$-$279, Dec. 2001.

\bibitem{ref_MM}
Y. Sun, P. Babu, and D. P. Palomar,
``Majorization-minimization algorithms in signal processing, communications, and machine learning,''
\emph{IEEE Trans. Signal Process.},
vol. 65, no. 3, pp. 794$-$816, Feb. 2017.

\bibitem{ref_Penalty_1}
J. Nocedal and S. Wright, 
\emph{Numerical Optimization}. 
New York, NY, USA:
Springer, 2006.

\bibitem{ref_Penalty_2}
X. Yu, D. Xu, Y. Sun, D. W. K. Ng, and R. Schober, 
``Robust and secure wireless communications via intelligent reflecting surfaces,'' 
\emph{IEEE J.  Sel. Areas  Commun.}, 
vol. 38, no. 11, pp. 2637$-$2652, Nov. 2020.


\bibitem{ref_CVX}
M. Grant and S. Boyd,
\emph{CVX: Matlab software for disciplined convex
programming}, version 2.1, http://cvxr.com/cvx, Mar. 2014.


\bibitem{ref_SDP_complexity}
M. S. Lobo, L. Vandenberghe, S. Boyd, and H. Lebret, 
``Applications of second-order cone programming,'' 
\emph{Linear Algebra its Appl.}, 
vol. 284, nos. 1-3, pp. 193$-$228, Nov. 1998.





\bibitem{ref_BCA}
D. P. Bertsekas,
``Nonlinear programming,''
\emph{Journal of the Operational Research Society},
vol. 48, no. 3, pp. 334-334, 1997.






\end{thebibliography}
\end{document}